\documentclass[a4paper,11pt]{article}
\pdfoutput=1 

\usepackage{jcappub} 

\usepackage[T1]{fontenc} 
\usepackage{graphicx}
\usepackage{caption}
\usepackage{amsmath}
\usepackage{multirow}
\usepackage{dirtytalk}
\usepackage{multirow}
\usepackage{subcaption}
\usepackage{floatrow}
\usepackage[dvipsnames]{xcolor}

\def\beq{\begin{equation}}
\def\eeq{\end{equation}}

\newcommand{\Mp}{M_\mathrm{Pl}}

\definecolor{verde}{rgb}{0,0.5,0}

\begin{document}

\title{Kinetic Gauge Friction in Natural Inflation} 

\author[a]{Martino
Michelotti,}
\author[a,b]{Rodrigo~Gonzalez~Quaglia,}
\author[a]{\qquad\qquad Ema Dimastrogiovanni,}
\author[c]{Matteo Fasiello,}
\author[a]{Diederik Roest}

\affiliation[a]{Van Swinderen Institute for Particle Physics and Gravity,
University of Groningen, Nijenborgh 3, 9747 AG Groningen, The Netherlands}
\affiliation[b]{Instituto de Ciencias F\'{\i}sicas, Universidad Nacional
Aut\'onoma de M\'exico,\\ Av. Universidad s/n, Cuernavaca, Morelos, 62210, Mexico}
\affiliation[c]{
Instituto de F\'isica T\'eorica UAM-CSIC, Calle Nicol\'as Cabrera 13-15, 28049, Madrid, Spain}

\emailAdd{m.michelotti@rug.nl}\emailAdd{r.gonzalez.quaglia@rug.nl}

\abstract{We study an extension of the natural inflation model comprising a non-Abelian gauge sector coupled to the axion-inflaton kinetic term. We show how such non-minimal coupling serves as a source of friction for the rolling inflaton granting sixty or more $e$-folds of accelerated expansion for sub-Planckian values of the axion decay constant. The analysis of perturbations reveals a negative sound speed, thus signaling an instability. Implementing a Chern-Simons-type coupling between the inflaton and gauge sectors cures the instability by delivering a positive speed. We perform a numerical study of scalar and tensor perturbations for a fiducial set of parameters finding that the corresponding observables are compatible with current CMB bounds.}

\maketitle
\flushbottom

\section{Introduction}\label{Introduction}
First conceived \cite{Guth:1980zm} as a solution to some of the puzzles of the hot big-bang cosmological model, inflation \cite{Linde:1984ir,Lyth:1998xn} has since become the most widely accepted paradigm for describing the dynamics of the very early universe. Besides offering an elegant solution to the so-called horizon problem, the inflationary mechanism provides the seeds required for the growth of structure in the universe. In its simpler realization, inflation, an early phase of accelerated expansion, is arrived at by means of a scalar field minimally coupled to gravity, whose potential must be sufficiently flat  to give rise to a quasi-de Sitter expansion. This is usually achieved in the so called single-field slow-roll regime. The potential ought to also ensure that inflation may end. There are countless \cite{Martin:2013tda} ways to implement a successful single-field inflationary mechanism (see \textit{e.g.} \cite{Linde:1983gd,Linde:1993cn,Kallosh:2013tua}). Naturally, multi-field models can also sustain such dynamics \cite{Starobinsky:2001xq,Wands:2007bd,Senatore:2010wk,Achucarro:2015rfa}, as can models with different gravitational theories \cite{Kannike:2015apa,Rinaldi:2015uvu,Starobinsky:1980te,DeFelice:2010aj} and geometries (see \textit{e.g.} \cite{Racioppi:2024zva,GonzalezQuaglia:2022kvs,Shaposhnikov:2020gts, Raatikainen:2019qey, Guimaraes:2020drj}). Nonetheless, in general, the flatness of the potential is difficult to maintain once radiative corrections\footnote{This notion can be equivalently formulated as the issue with quantum corrections giving the inflaton a large mass.} are taken into account. This is a well-studied issue in inflationary model building, known as the \textit{eta problem}, whose resolution can take different forms \cite{Freese:1990rb,Berera:2003yyp,Easson:2009kk,Ashoorioon:2011aa}.

\indent  One way to tackle the \textit{eta problem} is to employ a symmetry to ``protect'' the potential from large quantum corrections. A famous case in point is that of  natural inflation (NI)  \cite{Freese:1990rb,Adams:1992bn}, where a pseudo Nambu-Goldstone boson plays the role of the inflaton, whose cosine potential is the result of non-perturbative corrections\footnote{One can view the recently proposed modular inflation and $SL(2,Z)$ cosmology \cite{Schimmrigk, Casas:2024jbw, Kallosh:2024ymt} as extensions of this to the axion-dilaton sector.}.  This model, at least in its initial and simplest formulation, has recently been ruled  by the  Planck-BICEP/Keck Array data \cite{BICEP:2021xfz}. One should also note that the region of parameter space of natural inflation that has been known to perform best vis-\`{a}-vis CMB observations (see \textit{e.g.} \cite{Freese:2014nla}) corresponds to trans-Planckian axion decay constant values. This notion has to be paired up with the expectation that all global symmetries, including the shift symmetry, are broken at the Planck scale \cite{Holman:1992us}. This argument does not hold if the symmetry stems from a gauge symmetry as in string theory. Nevertheless, typical string theory constructions come with  a sub-Planckian axion decay constant $f$ \cite{Banks:2003sx}. 

One ought to see in this context the many NI-inspired efforts found in the literature to build a viable and (technically) natural inflationary model characterized by a sub-Planckian $f$. A rather common feature in these constructions is the presence of an additional source of friction besides the standard Hubble term. The resulting slowing-down of the rolling of the inflaton may sustain inflation even in the case of a relatively steep potential. In this sense adding friction serves the same purpose of increasing the value of the axion decay constant: one slows down the inflaton by resisting its downward motion, the other by merely reducing the slope of its descent. As we shall see, friction may be the result of non-minimal or non standard couplings of the inflaton with gravity or of a coupling with additional degrees of freedom. 

\indent An interesting proposal for a viable model in the natural inflation model class is the so called UV-protected NI \cite{Germani:2010hd,Germani:2011ua}. Here the kinetic term of the inflaton is non-minimally coupled to gravity, providing an additional source of friction (see also \cite{Dimastrogiovanni:2023oid}).
In the present work we will take inspiration from the UV-protected model and  consider the coupling of the kinetic term of the axion-inflaton with gauge fields 
in what we will refer to  as \textit{kinetic gauge coupling}.

\indent Another mechanism for friction stemming from the coupling of the rolling axion-like particle (ALP) with the gauge sector relies on the parity-breaking dimension-5 Chern-Simons term, although interesting alternatives exist \cite{Bartolo:2020gsh}. Such setup preserves the (approximate) shift symmetry of the NI Lagrangian and has received considerable attention both in the Abelian \cite{Anber:2009ua,Barnaby:2010vf,Barnaby:2011vw,Cook:2011hg,Mukohyama:2014gba,Ozsoy:2014sba,Namba:2015gja,Peloso:2016gqs,Garcia-Bellido:2016dkw,Ozsoy:2017blg,Domcke:2020zez,Ozsoy:2021onx,Talebian:2022jkb,Campeti:2022acx,Caravano:2022epk,Dimastrogiovanni:2023juq,Putti:2024uyr,Durrer:2024ibi,Alam:2024krt,Ozsoy:2024apn,Caravano:2024xsb} and non-Abelian \cite{Maleknejad:2011sq,Maleknejad:2011jw,Adshead:2012kp,Dimastrogiovanni:2012st,Dimastrogiovanni:2012ew,Adshead:2013qp,Adshead:2013nka,Namba:2013kia,Obata:2016tmo,Dimastrogiovanni:2016fuu,Adshead:2016omu,Caldwell:2017chz,Agrawal:2017awz,Thorne:2017jft,Lozanov:2018kpk,Dimastrogiovanni:2018xnn,Domcke:2018rvv,Fujita:2018vmv,DallAgata:2018ybl,Mirzagholi:2020irt,Watanabe:2020ctz,Campeti:2020xwn,Holland:2020jdh,Iarygina:2021bxq,Fujita:2021flu,Ishiwata:2021yne,Adshead:2022ecl,Fujita:2022fff,Bagherian:2022mau,Iarygina:2023mtj,Dimastrogiovanni:2024xvc,Dimastrogiovanni:2024lzj,Badger:2024ekb,Murata:2024urv,Brandenburg:2024awd} cases. One model of particular interest for us is the 
chromo-natural Lagrangian of \cite{Adshead:2012kp} whose rich gravitational wave (GW) phenomenology has received a great deal of scrutiny  \cite{Dimastrogiovanni:2016fuu,Adshead:2013qp,Ozsoy:2024apn,Obata:2016tmo}, also in view of the opportunity to test these ideas with upcoming GW probes, from pulsar timing arrays to interferometers.\\

\noindent This paper is organized as follows. In Section \ref{Brief review of Natural Inflation and friction mechanisms } we review the natural inflation proposal, together with two of its extensions, UV-protected natural inflation and chromo-natural inflation (CNI). In Section \ref{Section background} we add a non-Abelian gauge sector to the NI Lagrangian by coupling gauge fields to the kinetic term of the axion-inflaton, implementing what we call \textit{kinetic gauge friction}. We solve the background equations of this model and explicitly show how a sufficiently long period of inflation can be obtained. In Section \ref{Kinetic gauge friction: Perturbations} we present the analysis of tensor perturbations and show how a Chern-Simons (CS) term may be introduced to cure an instability of the modes. We call this setup, which combines kinetic gauge friction and the CS term, \textit{kinetic gauge friction chromo-natural} (KFC) inflation. In Section \ref{Kinetic gauge friction: Scalar perturbations} we study scalar perturbations finding that the vacuum of the scalar modes does not correspond to the standard Bunch-Davies solution. We then analyse in detail the initial conditions for scalar perturbations together with the phenomenological implications at the level of the power spectra. Finally, conclusions and future work are discussed in Section \ref{Conclusions}. Additional details on the background equations of motion and on fluctuations can be found in the Appendices.

\section{Brief review of Natural Inflation and friction mechanisms }\label{Brief review of Natural Inflation and friction mechanisms }
\subsection{Natural Inflation}
The Natural Inflation (NI) model \cite{Freese:1990rb,Adams:1992bn} is characterized by a minimally coupled inflaton field plus gravity; the action reads
\begin{equation}\label{Natural Inflation}
    S=\int d^{4}x\sqrt{-g}\left[\frac{\Mp^2}{2}R-\frac{1}{2}g^{\mu\nu}\partial_{\mu}\chi\partial_{\nu}\chi-V(\chi)\right],
\end{equation}
where the potential is given by
\begin{equation}\label{Natural Inflation Potential}
    V(\chi)=\mu^{4}\left(1+\cos\left
    (\frac{\chi}{f}\right)\right).
\end{equation}
Here $R$ is the Ricci scalar, $g$ is the determinant of the metric, $\mu$ is the scale of the potential, $\chi$ is the pseudo Nambu-Goldstone boson field responsible for inflation and $f$ its decay constant. 

\noindent The equations of motion (EoM) in a flat FRLW background are

\begin{equation}\label{NaturalInflationEoM}
    \ddot{\chi}+3H\dot{\chi}+V_{\chi}=0\,, \quad 3\Mp^{2}H^{2}=\frac{1}{2}\dot{\chi}^{2}+V(\chi)\,.
\end{equation}
Throughout the paper, the sub-index $(_{\chi})$ indicates a derivative w.r.t the field $\chi$ while an overdot implies derivative w.r.t cosmic time $t$. The Hubble parameter is defined as $H\equiv\dot{a}/a$.  The slow-roll limit, where the motion of the inflaton field is overdamped $(\ddot{\chi}\ll H\dot{\chi})$, is characterized by the following two conditions on the slow-roll parameters: 
\begin{equation}\label{slowrollparaV}
    \epsilon_{V}\equiv \frac{\Mp^2}{2}\left(\frac{V_{\chi}}{V}\right)^{2} \ll 1\,,\quad\eta_{V}\equiv\Mp^2\frac{V_{\chi\chi}}{V} \ll 1\,.
\end{equation}
These ensure that the potential has a flat direction and that it stays (nearly) flat for  the amount of time necessary to guarantee a sufficient duration of inflation. The inflationary observables $r, n_s$ (tensor-to-scalar ratio and scalar spectral tilt respectively) are related to the slow-roll parameters as follows: 
\begin{equation}
    r=16\epsilon_{V}\,, \quad
     n_{s}=1-6\epsilon_{V}+2\eta_{V}\,.\label{epsilonr-ns} 
\end{equation}
From Eqs.~\eqref{Natural Inflation Potential}, \eqref{slowrollparaV} and (\ref{epsilonr-ns})  one can readily obtain \cite{German:2021jer}
\begin{equation}\label{chiNI}
    \cos\left(\frac{\chi}{f}\right)=\frac{16\Mp^{2}}{8\Mp^{2}+rf^{2}}-1\,.
\end{equation}
From here one finds
\begin{equation}
    f=\frac{2}{\sqrt{4(1-n_{s})-r}}\Mp\,.
    \label{fobs}
\end{equation}
It is straightforward to verify through Eq.~(\ref{fobs}) that only trans-Planckian values for $f$ are compatible with current experimental bounds in the $(r,n_s)$ plane. Nevertheless,
upon requiring a minimum duration of inflation of about sixty $e$-folds, one is led to conclude that the simplest formulation of natural inflation is ruled out by observations \cite{Planck:2018jri,BICEP:2021xfz}. 

One proven way to alleviate such tension with the data is to incorporate an additional source of friction. The resulting slowing-down of the ALP down its potential may indeed leave room for a parameter space where a sub-Planckian decay constant is compatible with observations.

\subsection{UV-protected Natural Inflation}\label{Gravitatioaly enhanced Chromo Natural Inflation}
Perhaps the simplest way to add friction to the standard Hubble term ($\sim 3H \dot{\chi}$)  
is to contract the kinetic term of the axion-inflaton $\chi$
(also) with the Einstein tensor $G^{\mu\nu}$, to obtain what is known as the UV-protected natural inflation (UVNI) model \cite{Germani:2010hd,Germani:2011ua}:
\begin{equation}\label{GCNI action}
    S=\int d^{4}x\sqrt{-g}\left[\frac{\Mp^2}{2}R-\frac{1}{2}\left(g^{\mu\nu}-\frac{G^{\mu\nu}}{M^{2}}\right)\partial_{\mu}\chi\partial_{\nu}\chi-V(\chi)\right]\,.
\end{equation}
The specific non-minimal coupling employed here guarantees  second order EoM\footnote{One should add that there exists a vast literature on modified gravity models that nevertheless deliver effectively second-order equations of motion, from Horndeski \cite{Horndeski:1974wa} to DHOST \cite{Langlois:2015cwa,BenAchour:2016cay} theories. These ideas have also been explored in the natural inflation context, see \textit{e.g.} \cite{Kobayashi:2010cm,Burrage:2010cu,Watanabe:2020ctz,Murata:2024urv}.}
 while in general other options such as $R^{\mu\nu}\partial_{\mu}\chi\partial_{\nu}\chi$ suffer from ghost modes and instabilities \cite{Wald:1984rg}. 

 The inflaton EoM coming from Eq.~\eqref{GCNI action} has the following form:
\begin{equation}\label{GCNI EoM}
    \ddot{\chi}\left(1+\frac{3H^{2}}{M^{2}}\right)+3H\dot{\chi}\left(1+\frac{3H^{2}}{M^{2}}+\frac{2\dot{H}}{M^{2}}\right)-\frac{\mu^{4}}{f}\sin\left(\frac{\chi}{f}\right)=0\,.
\end{equation}
One can immediately notice that the new term ($\propto 1/M^{2}$) contributes to friction and that  the standard single-field slow-roll EoM  is recovered in the large $M$ limit. \\

\subsection{Chromo-Natural Inflation}

A more radical path to a friction term is arrived at by coupling the axion-inflaton to additional degrees of freedom. The most studied scenario is that of a Chern-Simons (CS) type coupling to a gauge sector that serves as a dissipation channel for the inflaton kinetic energy. Both in the Abelian and non-Abelian cases, gauge fields provide a friction term in the inflaton EoM.\footnote{In the Abelian case, the gauge field VEV is often set to zero in light of stringent bounds on anisotropy. The friction term is then proportional to fluctuations in the gauge sector. The non-Abelian scenario, starting with the SU(2) configuration, may instead feature a non-zero VEV whilst maintaining background isotropy \cite{Maleknejad:2011jw}.} The CS coupling does not break (up to a total derivative) the shift symmetry and as such preserves the naturalness of its NI progenitor. There is of course a price to pay with respect to the UVNI model, namely the extra degrees of freedom in the gauge sector. On the other hand, these give rise to a very interesting gravitational wave phenomenology spanning from a non-trivial primordial GW spectral shape to non-zero GW chirality \cite{Thorne:2017jft,Campeti:2020xwn}, from large GW non-Gaussianities \cite{Agrawal:2017awz,Dimastrogiovanni:2018xnn,Fujita:2018vmv} to primordial black hole production \cite{Garcia-Bellido:2016dkw,Garcia-Bellido:2017aan,Dimastrogiovanni:2024xvc}.

We shall focus here on a non-Abelian SU(2) gauge sector describing the chromo-natural inflation model \cite{Adshead:2012kp}:
\begin{equation}\label{Chromo Natural Inflation}
    S=\int d^{4}x\sqrt{-g}\left[\frac{\Mp^2}{2}R-\frac{1}{2}g^{\mu\nu}\partial_{\mu}\chi\partial_{\nu}\chi-V(\chi)-\frac{1}{4}F^{a\mu\nu}F^{a}_{\mu\nu}-\frac{\lambda\chi}{8f\sqrt{-g}}\epsilon^{\mu\nu\sigma\rho} F^{a}_{\mu\nu} F^{a}_{\rho\sigma}\right]\,,
\end{equation}
where $F^{a}_{\mu\nu}$ is the SU$(2)$ field strength of $A^a_{\mu}$ defined as

\begin{equation}
    F^{a}_{\mu\nu}=\partial_{\mu}A^{a}_{\nu}-\partial_{\nu}A^{a}_{\mu}+g\epsilon^{abc}A^{b}_{\mu}A^{c}_{\nu}\,.
\end{equation}
Here $g$ is the gauge field self interaction coupling constant, not to be confused with the determinant of the metric, while $\lambda$ is the dimensionless coupling between the inflaton and the gauge fields. Greek indices correspond to spacetime indices, while Roman ones indicate the gauge indices. Our convention for the totally antisymmetric symbol is $\epsilon^{0123}=1$.
The CNI prescription for the VEV is as follows:
\begin{equation}\label{Gauge field decomposition}
    A_{0}^{a}=0\,, \quad A_{i}^{a}=\delta_{i}^{a}a(t)Q(t)\,,
\end{equation}
with $Q(t)$ being the gauge field vacuum expectation value (vev). It is important to stress here that the isotropic one is an attractor solution \cite{Maleknejad:2013npa,Wolfson:2021fya} and that, even when starting out with a zero vev, a non-zero one will dynamically develop in this scenario \cite{Domcke:2018rvv}.

The  Chern-Simons interaction in CNI (Eq.~(\ref{Chromo Natural Inflation})) enables an energy transfer from the inflaton to  gauge fields, effectively slowing down the rolling of the scalar. The inflaton EoM reads 
\begin{equation}\label{CNI EoM}
    \ddot{\chi}+3H\dot{\chi}-\frac{\mu^{4}}{f}\sin\left(\frac{\chi}{f}\right)=-\frac{3g\lambda}{f}Q^{2}\left(\dot{Q}+HQ\right)\,.
\end{equation}
It is also useful to introduce the particle production parameter $\xi\equiv \lambda \dot{\chi}/(2fH)$. Qualitatively, one can understand the effect of the Chern-Simons term as follows. Early on in the dynamics, when the ALP is rolling down the flatter part of the potential, $\chi$ is almost constant (and therefore $\xi$ is very small) and the Chern-Simons interaction in Eq.~\eqref{Chromo Natural Inflation} is well approximated by a boundary term, effectively decoupling from the scalar field dynamics so that the evolution reduces to that in NI. As the inflaton velocity (and $\xi$) increases, the interaction with the gauge sector starts having an impact, with the inflaton damping energy into the gauge sector as can be seen by the source term in Eq.~(\ref{CNI EoM}). 

Following the same logic, at the level of perturbations, the parameter $\xi$ is indeed the best indicator of gauge quanta production. This mechanism for particle  production turns out to be extremely effective.  Given its growth as a function of the inflaton velocity, it is intuitively clear that fluctuations in the gauge sector may increase to the point 
that their backreaction on the background EoM is no longer negligible. This is the so called strong backreaction regime, currently a topic of intense research activity by means of analytical and numerical  techniques \cite{Anber:2009ua,Dimastrogiovanni:2016fuu,Peloso:2022ovc,Garcia-Bellido:2023ser,Iarygina:2023mtj,Dimastrogiovanni:2024xvc}, as well as  lattice simulations \cite{Caravano:2022epk,Figueroa:2023oxc,Caravano:2024xsb}.\\
\indent The CNI setup  has been shown to alleviate the tension with CMB data with respect to its NI counterpart and to be able to do so for sub-Planckian $f$ values. It is nevertheless ruled out by the most recent CMB observations \cite{Planck:2018jri,BICEP:2021xfz}. CNI remains a very widely studied model due to its very rich phenomenology. Indeed, one should  point out here that CNI extensions such as  Higgs-ed CNI \cite{Adshead:2016omu}, spectator CNI \cite{Dimastrogiovanni:2016fuu} and non-minimally coupled CNI \cite{Dimastrogiovanni:2023oid} are cosmologically viable models that share with CNI its distinct and, most importantly, testable signatures.

\section{Background evolution}\label{Section background}
\subsection{Kinetic gauge friction}

\noindent In this paper, we propose an alternative source of  friction, which we call \textit{kinetic gauge friction}, analogous to that in Eq.~\eqref{GCNI action} but with the role of the Einstein tensor being now  played by non-Abelian gauge fields:

\begin{equation}\label{GaugeAction}
    \hspace{-0.5cm}S=\int d^{4}x\sqrt{-g}\left(\frac{\Mp^{2}}{2}R-\frac{1}{2}\left(g^{\mu\nu}+\lambda_{1}F^{a\mu}_{\ \ \ \alpha}F^{a\alpha\nu}+\lambda_{2}F^{a\rho\sigma}F^{a}_{\ \rho\sigma}g^{\mu\nu}\right)\partial_{\mu}\chi\partial_{\nu}\chi-V(\chi)-\frac{1}{4}F^{a}_{\mu\nu}F_{a}^{\mu\nu}\right)\,.
\end{equation}
Here $V(\chi)$ is the NI  potential in Eq.~\eqref{Natural Inflation Potential}; the coupling constants $\lambda_{1,2}$ have units of $[E^{-4}]$, given that the new operators are dimension $8$. 
 From the EFT perspective the novel terms in Eq.~(\ref{GaugeAction}) can be thought of as accounting for the next-to-leading order corrections  to the Lagrangian that are both shift symmetric and parity-even in the kinetic sector.
 
\indent The system of background EoM for generic $\lambda_{1}$ and $\lambda_{2}$ couplings can be found  in Appendix \ref{appA}. In the remainder of the main text we  focus on the special case $\lambda_{2}=\lambda_{1}/2$, which leads to a simplification of the system of equations. 
The Friedmann, axion and gauge field background EoM then read
\begin{equation}\label{KGFHEoM}
    3\Mp^{2}H^2=\frac{1}{2} \dot{\chi}^2 \Big(1+3g^{2}\lambda_{1}Q^{4}\Big)+\frac{3}{2} \left(H Q+\dot{Q}\right)^{2}+\frac{3}{2} g^{2} Q^{4}+V(\chi)\,,
\end{equation}
\begin{equation}\label{KGFHPEoM}
    -2\Mp^{2}\dot{H}=\dot{\chi}^{2}\left(1+g^{2} \lambda_{1}Q^{4}\right)+2 \left(H Q+\dot{Q}\right)^{2}+2 g^{2} Q^{4}\,,
\end{equation}
\begin{equation}\label{KGFXEoM}
    \ddot{\chi}\left(1+3g^{2}\lambda_{1}Q^{4}\right)+3H\dot{\chi} \left(1+\frac{4g^{2}\lambda_{1}Q^{3} \dot{Q}}{H}+3g^{2} \lambda_{1}Q^4\right)+V_{\chi}(\chi)=0\,,
\end{equation}

\begin{equation}\label{KGFQEoM}
    \ddot{Q}+3H\dot{Q}+Q\left(\dot{H}+2 H^{2}\right)+2g^{2}Q^{3}=2g^{2}\lambda_{1} Q^{3}\dot{\chi}^{2}\,.
\end{equation}
The new source of friction in the axion equation depends both on the sign of $\lambda_{1}$ and on the evolution of the background quantities. As long as $\lambda_{1}$ is positive, the kinetic gauge coupling will act as a source of friction modulated by the combination $g^{2}\lambda_{1}$. Notice that setting $\lambda_2=\lambda_1/2$ leads to the cancellation of all the $\lambda_{1,2}$ terms that are not proportional to $g$, thus leaving only non-Abelian contributions to the friction.\\
\indent We plot in the left panel of Fig.~\ref{relative magnitudes} the magnitude of the kinetic gauge friction (neglecting $\dot{Q}$ terms, see subsection \ref{slowroll}) and the Hubble friction, normalized to the derivative of the axion potential, where it is clear that the former is dominating throughout the whole inflationary period.
For this and the remaining figures in the manuscript we use the following fiducial set of values for the model parameters (we discuss our choice of parameters in Sec.~\ref{Observables}):
\begin{equation}\label{benchmark}
\{ g= 0.5, \,\lambda_{1}= \left(1\times 10^{4}\Mp^{-1}\right)^{4},\, \mu=2.2\times10^{-3}\Mp, \,f= 0.24\Mp \}\,.
\end{equation}
While in Fig.~\ref{relative magnitudes} the kinetic gauge friction dominates over Hubble friction, this is not necessarily the case for other choices of parameters. \\
\indent One may wonder how natural is the value set for the kinetic gauge friction coupling constant $\lambda_1$. Considering $\lambda_1^{-1/4}=10^{-4}\Mp$, we can compare it with the mass scale of the gravitationally-enhanced frictions in previous works \cite{Germani:2011ua,Dimastrogiovanni:2023oid}, where $M\simeq10^{-8}\Mp$ and $M\simeq10^{-6}\Mp$, respectively. This shows that the value of $\lambda_1$ required for the kinetic gauge friction to be important is not far from typical values of mass-dimensionful couplings in the literature. \\

\subsection{Slow-roll regime}\label{slowroll}

\begin{figure}[t!]
    \includegraphics[scale=0.35]{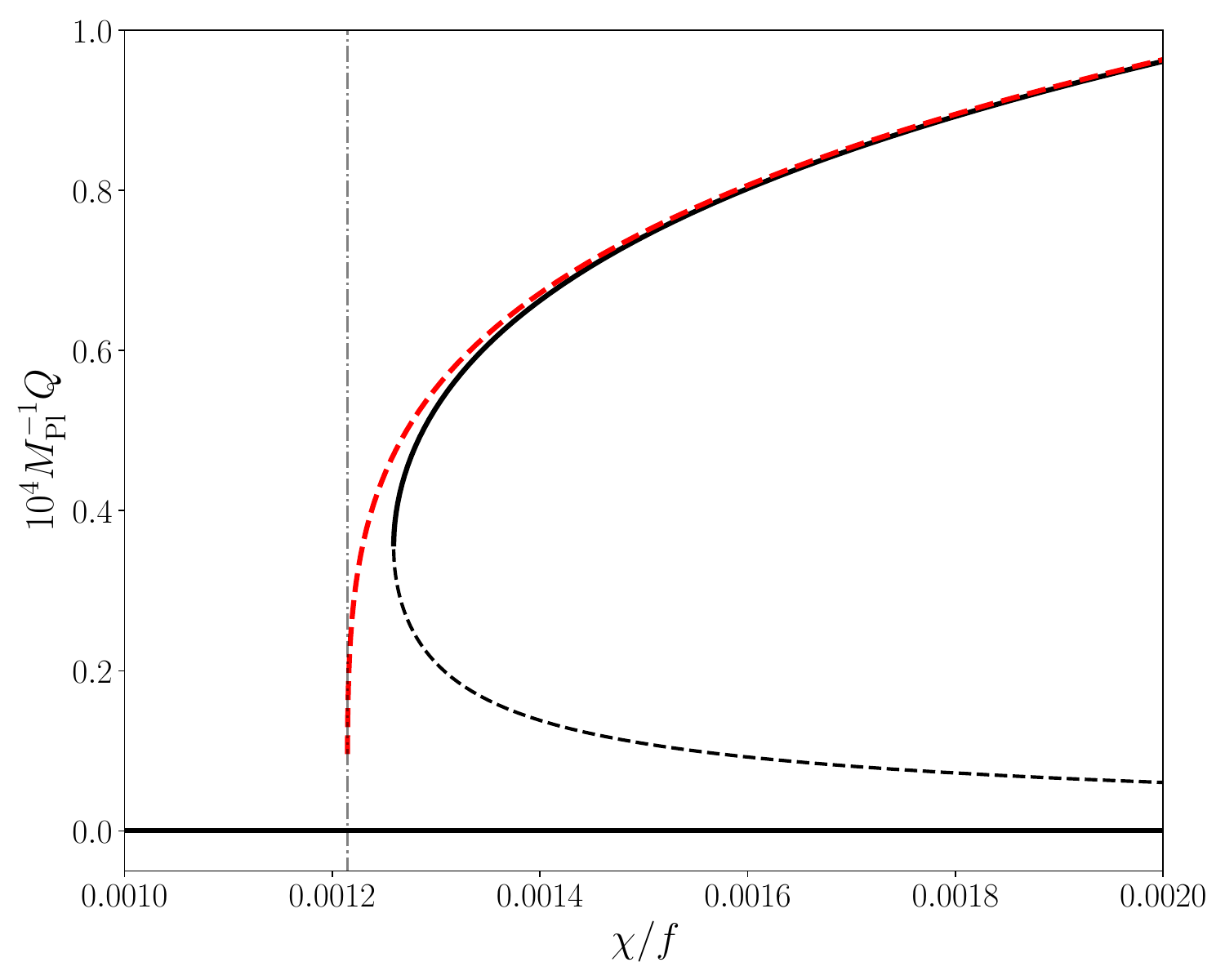}
    \caption{\textit{Plot depicting all the (non-negative) real solutions for $Q$; those corresponding to the minima are in solid black while the dotted black line corresponds to the maximum. The analytical approximation given in Eq.~(\ref{Corrected Q}) is shown in dashed red. This configuration is mirrored for negative values of $Q$. The vertical dash-dotted line indicates the analytical estimate for the value of $\chi_{\rm min}$ in Eq.~(\ref{chimin}), where non-zero values of $Q$ develop for $\chi>\chi_{\rm min}$.}}\label{FullSolutionsQ}
\end{figure}

 Under standard slow-roll assumptions, we drop every acceleration term in Eqs.~\eqref{KGFHEoM}-\eqref{KGFQEoM} and consider the Hubble parameter and the gauge field vev to be  slowly varying ($\dot{Q}\ll HQ$). Given our focus on the region of parameter space where the kinetic gauge friction dominates over the Hubble one, we require that  $1 \ll 3g^2\lambda_1Q^4$. Yet another assumption takes the form $H\ll gQ$. The role of the last inequality is making sure we stay clear from the well-known instability in the scalar sector encountered as soon as $\sqrt{2}H>gQ$ \cite{Dimastrogiovanni:2012ew}. In this regime, the Friedmann equation \eqref{KGFHEoM} reduces to that of single scalar field inflation (see also the right panel of Fig.~$\ref{relative magnitudes}$):
\begin{figure}[t]
    \begin{subfigure}[b]{0.45\textwidth}
        \centering
        \includegraphics[scale=0.25]{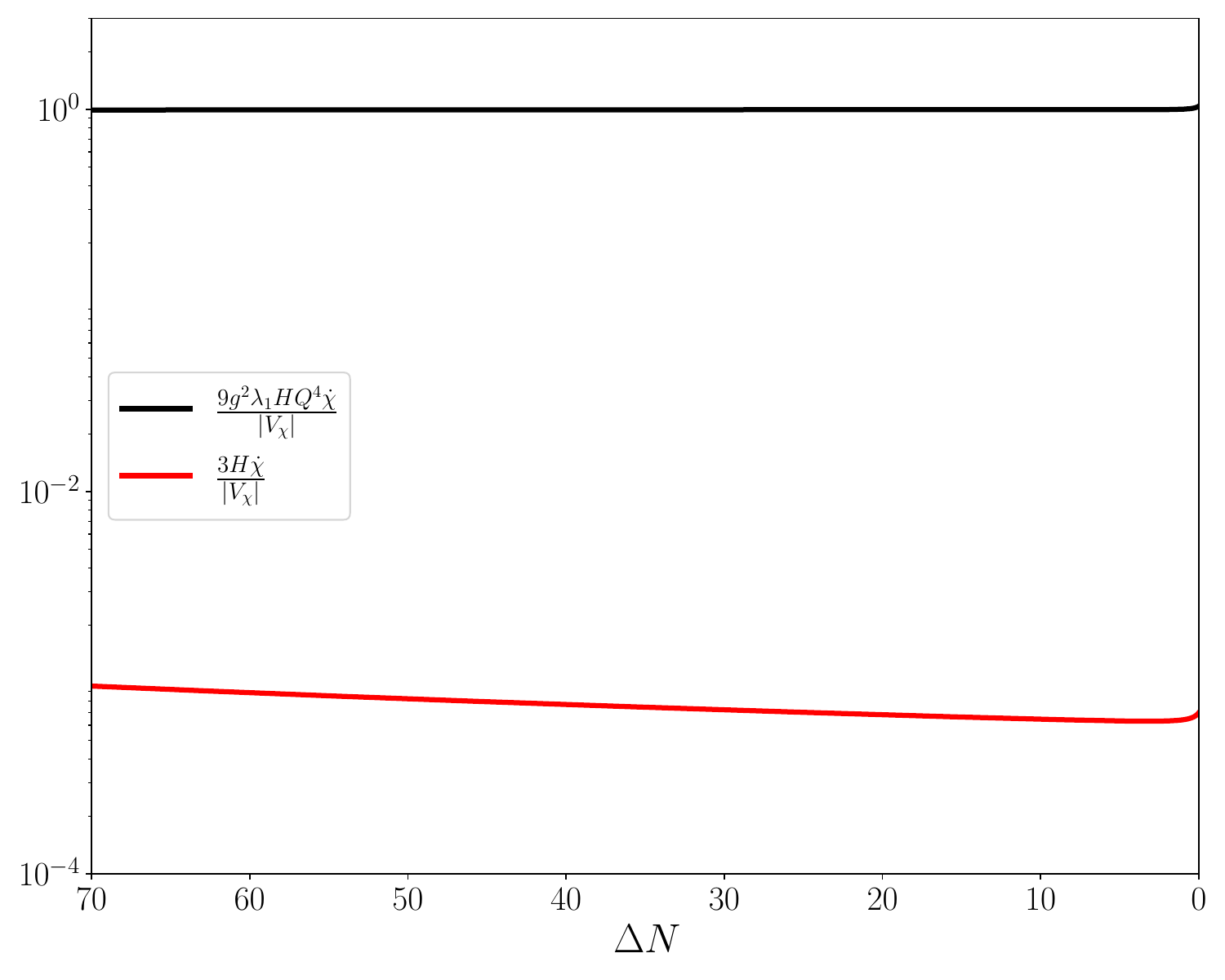}
    \end{subfigure}
    \hfill
    \begin{subfigure}[b]{0.45\textwidth}
        \centering
        \includegraphics[scale=0.25]{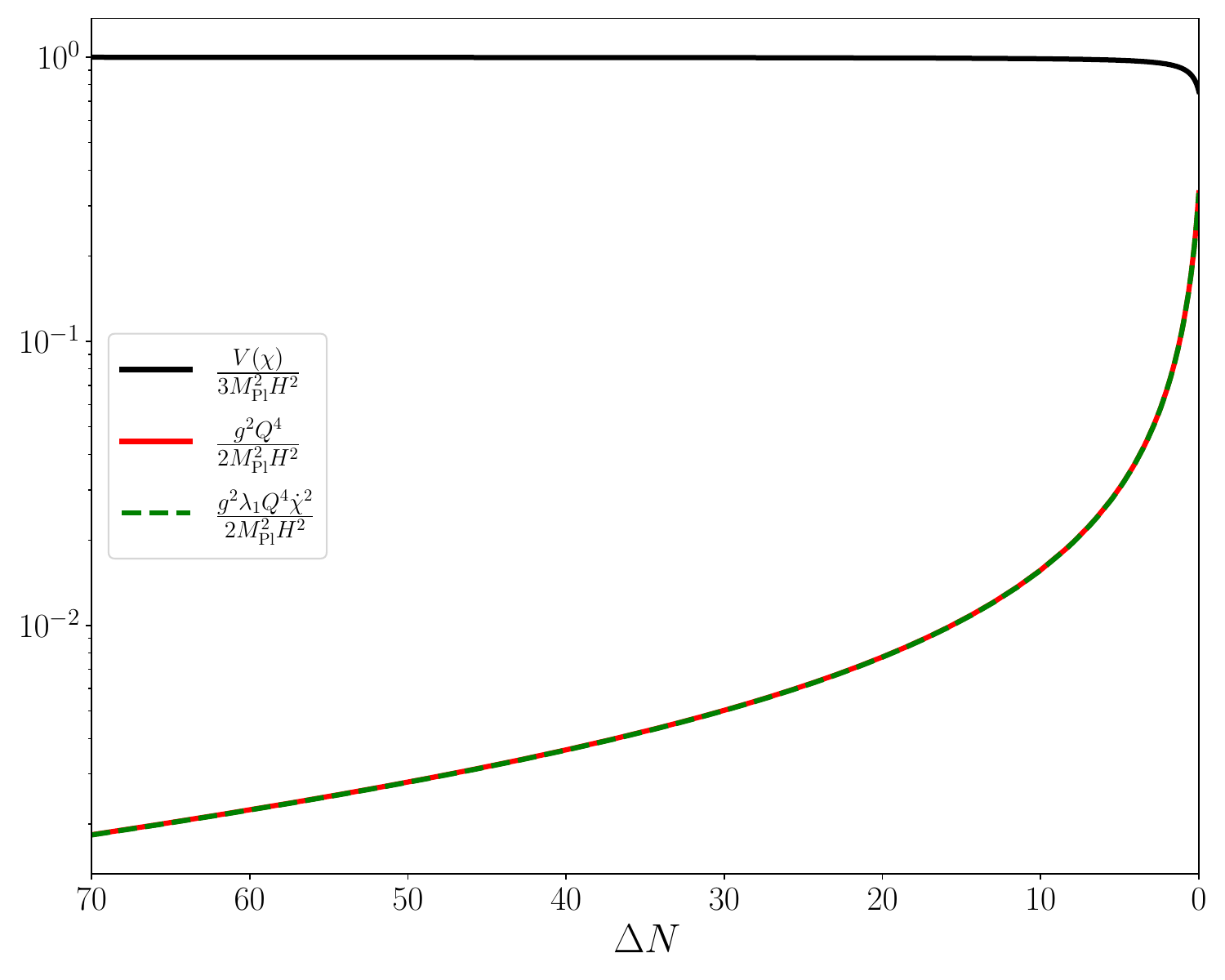}
    \end{subfigure}
    \caption{\textit{Left panel: Plot of the relative magnitude of the leading friction terms in Eq.~(\ref{KGFXEoM}). Right panel: Plot of the relative magnitude of the leading contributions to the energy density in  Eq.~(\ref{KGFHEoM}). In these and the following plots we use as time variable the number of $e$-folds $\Delta N$ from the end of inflation ($N_{\rm end}=0$). Both plots were generated  employing the fiducial set of parameters (Eq.~(\ref{benchmark}))}.}
    \label{relative magnitudes}
\end{figure}
\begin{equation}\label{SlowRollPot}
    3\Mp^{2}H^{2}\simeq V(\chi)\,,
\end{equation}
while the axion and gauge field equations take the following form:
\begin{equation}\label{SlowRollX}
    \dot{\chi}\simeq-\frac{V_{\chi}(\chi)}{9H g^{2} \lambda_{1} Q^{4}}\,,\quad\quad\dot{Q}\simeq-\frac{2g^{2}Q^{3} \left(1-\lambda_{1} \dot{\chi}^{2}\right)}{3H}\,.
\end{equation}
The $Q$ background value is found by setting $\dot{Q}=0$ in the second equation, which results in
\begin{equation}\label{SlowRollChidot}
    \dot{\chi}\simeq\frac{1}{\sqrt{\lambda_1}}\,,
\end{equation}
where we have disregarded the trivial solution $Q=0$, as it would lead to no contribution of the gauge field at the background level. This then allows us to find a solution for the gauge field vev: 

\begin{equation}\label{SlowRollQ}
    Q\simeq\left(-\frac{V_{\chi}(\chi)}{9g^{2}\sqrt{\lambda_{1}}H}\right)^{1/4}\,.
\end{equation}
It is convenient to use the number of $e$-folds
\begin{equation}\label{efoldsdef}
    dN=Hdt=\frac{H}{\dot{\chi}}d\chi\,,
\end{equation}
as the independent time variable to solve analytically for the scalar and gauge fields evolution. Substituting Eq.~\eqref{SlowRollPot} and Eq.~\eqref{SlowRollChidot} into Eq.~\eqref{efoldsdef} one first finds the following relation: 
\begin{equation}\label{e folds}
    \Delta N\equiv N_{\rm end}-N\simeq\frac{\mu^{2}\sqrt{\lambda_{1}}}{\sqrt{3}\Mp}\int_{\chi}^{\chi_{\rm end}}\sqrt{1+\cos\left(\frac{\chi}{f}\right)}d\chi\,.
\end{equation}
Upon setting $\chi_{\rm end}=\pi f$, Eq.~(\ref{e folds}) can  be integrated and inverted to obtain\footnote{Note that, in order to recover the NI regime as $\chi$ approaches the origin, one would 
need to reinstate the Hubble friction term in Eq.~\eqref{SlowRollX} and eventually Eq.~(\ref{ChiN}). }
\begin{equation}\label{ChiN}
    \chi(\Delta N)\simeq2f\arcsin\left(1-\frac{\sqrt{3}\Mp}{2\sqrt{2\lambda_1}f\mu^2}\Delta N\right)\,.
\end{equation}
Plugging Eq.~\eqref{ChiN} back into Eq.~\eqref{SlowRollPot} one gets
\begin{equation}\label{HN}
    H(\Delta N)\simeq\sqrt{\frac{2}{3}}\frac{\mu^{2}}{\Mp}\sqrt{1-\left(1-\frac{\sqrt{3}\Mp}{2\sqrt{2\lambda_{1}}f\mu^{2}}\Delta N\right)^{2}}\,,
\end{equation}
and similarly, from Eq.~\eqref{SlowRollQ}, one can obtain an expression for the vev of the gauge field:
\begin{equation}\label{QN}
    Q(\Delta N)\simeq\left(\sqrt{\frac{2}{27\lambda_{1}}}\frac{\mu^{2}\Mp}{fg^{2}}\left(1-\frac{\sqrt{3}\Mp}{2\sqrt{2\lambda_{1}}f\mu^2}\Delta N\right)\right)^{1/4}\,,
\end{equation}
where the trigonometric identities
\begin{equation*}
    \sin(2\arcsin x)=2x\sqrt{1-x^2}\,, \quad \cos(2\arcsin x)=1-2x^2\,,
\end{equation*}
were employed. All three quantities of interest, the axion and gauge field background values, together with the Hubble parameter, determine the background evolution of the system. Here we also report a 
formula for the slow-roll parameter $\epsilon_H\equiv-\dot{H}/H^{2}$, given its importance to determine the end of inflation:

\begin{equation}\label{epsilonN}
    \epsilon_H(\Delta N)\simeq\frac{1}{2}\left(\frac{1}{\Delta N}+\frac{3\Mp}{3\Mp\Delta N-4\sqrt{6\lambda_1}f\mu^2}\right)\, .
\end{equation}
The analytical solutions for $Q$ and $\epsilon_{H}$ are plotted in Fig.~\ref{fig:comparison_lambda} together with their full numerical solutions, showing a good agreement between the two results. We verified that this is also the case for $\chi$ and $H$.\\
\indent We will now derive an expression for $\Delta N$ as a function of the model parameters and show that it is possible to achieve over sixty $e$-folds of inflation without resorting to super-Planckian $f$ values. To this end, we first obtain a more precise analytic solution for $Q$ by relaxing the assumption $1 \ll 3g^2\lambda_1Q^4$ in Eq.~(\ref{KGFQEoM})\footnote{Numerically, one can verify that Eq.~(\ref{KGFQEoM}) has five real (symmetric) solutions for $Q$, the trivial one ($Q=0$), two positive ones (one corresponding to a minimum and one to a maximum of the effective potential) and their negative counterparts. Of these solutions, Eq.~(\ref{Corrected Q}) is the analytic approximation to the positive minimum. All non-negative solutions are plotted in Fig.~\ref{FullSolutionsQ}.}, which becomes 
\begin{equation}\label{Corrected Q}
    Q\simeq\left(-\frac{\sqrt{\lambda_{1}}V_\chi+3H}{9g^{2}\lambda_{1}H}\right)^{1/4}\,.
\end{equation}
In Fig.~\ref{FullSolutionsQ} we show the full numerical solutions for $Q$ (in the slow-roll approximation) together with the analytical estimate given in Eq.~\eqref{Corrected Q}. We now derive the smallest  value of $\chi_{\text{in}}$ compatible with  a positive  solution for $Q$. Setting the right-hand-side of Eq.~\eqref{Corrected Q} to be positive and using Eq.~\eqref{SlowRollPot} one finds

\begin{figure}[t!]
    \centering
    \includegraphics[scale=0.35]{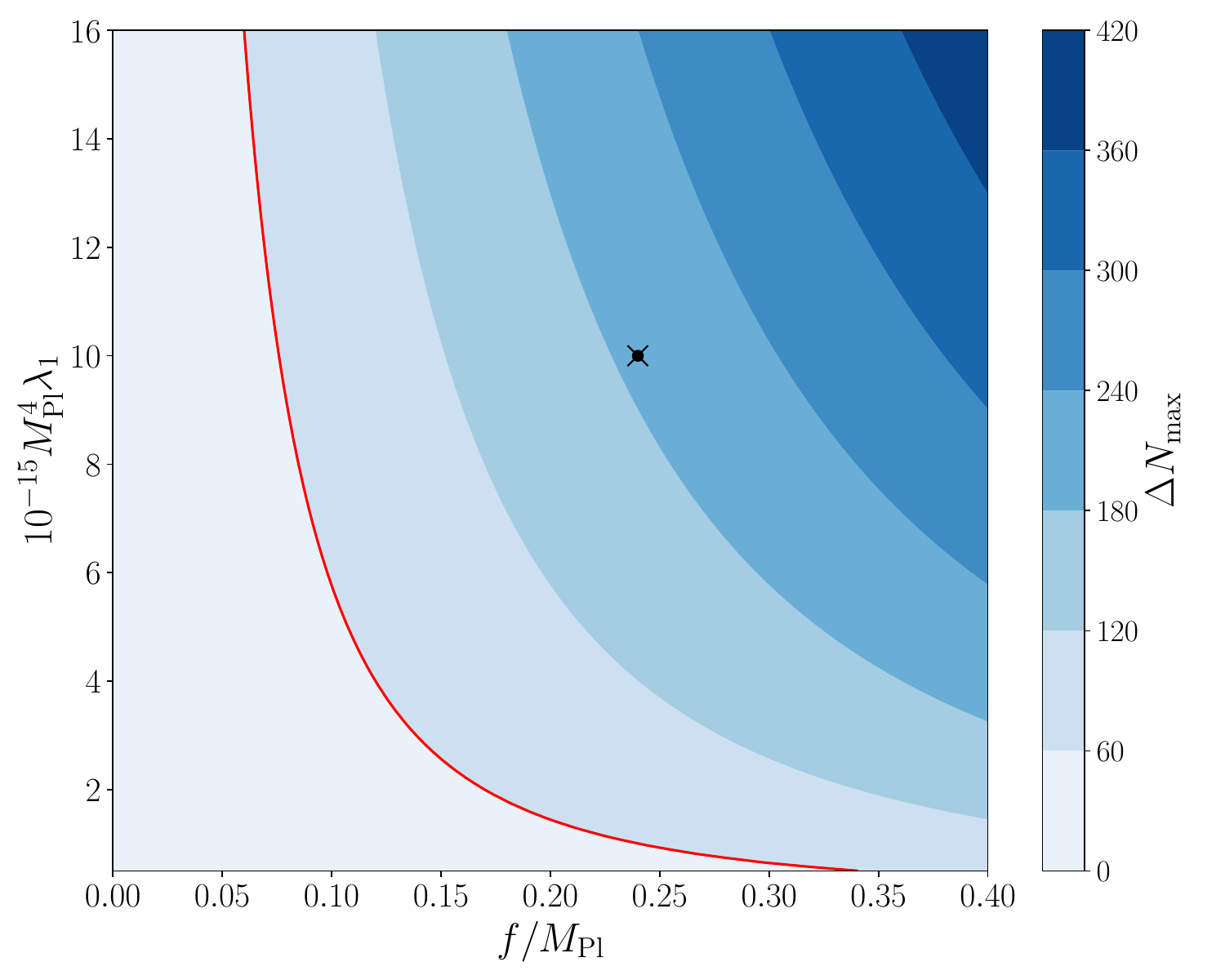}
    \caption{\textit{Contourplot of the maximum duration of inflation, $\Delta N_{\rm max}$, as a function of $f$ and $\lambda_1$, where the remaining parameters are set to the fiducial values in Eq.~(\ref{benchmark}). The plot shows that, generically, a period of inflation that lasts at least sixty $e$-folds (red line) is achieved for sub-Planckian values of the decay constant $f$ as long as $\lambda_{1}\gtrsim10^{15}\Mp^{-4}$.The black cross corresponds to the fiducial set of parameters.}}
    \label{Delta N vs parameters}
\end{figure}

\begin{equation}\label{chimin}
    \chi_{\rm min}\simeq\frac{\sqrt{6}f^2}{\sqrt{\lambda_{1}}\mu^2\Mp}\,,
\end{equation}
where we are assuming $\chi_{\rm min}/f\ll1$. With this relation at hand, it is  possible to compute the maximum amount of $e$-folds as a function of the model parameters:
\begin{equation}\label{Nfbounds}
   \Delta N_{\rm max}\simeq \frac{2\sqrt{2\lambda_{1}}f\mu^{2}}{\sqrt{3} \Mp}\left(1-\sin \left(\frac{\sqrt{3}f}{\sqrt{2\lambda_{1}}\mu ^2\Mp}\right)\right)\,.
\end{equation}
In Fig. \ref{Delta N vs parameters} we show the parameter space $\{f,\lambda_1\}$ compatible with $\Delta N_{\rm max}\geq60$.

\section{Tensor Perturbations}\label{Kinetic gauge friction: Perturbations}

Having explored the effect of the kinetic gauge coupling at the background level, we now move to study the perturbations. We start with tensors and leave scalars to the next section. Vector modes are expected to decay in this and similar setups \cite{Adshead:2013nka}. 
We  first take on  tensor fluctuations in the parity-even model and later add the effect of the CS term.

\subsection{Tensor perturbations decomposition}\label{Tensor perturbations decomposition}

\begin{figure}[t]
    \begin{subfigure}[b]{0.45\textwidth}
        \centering
        \includegraphics[scale=0.25]{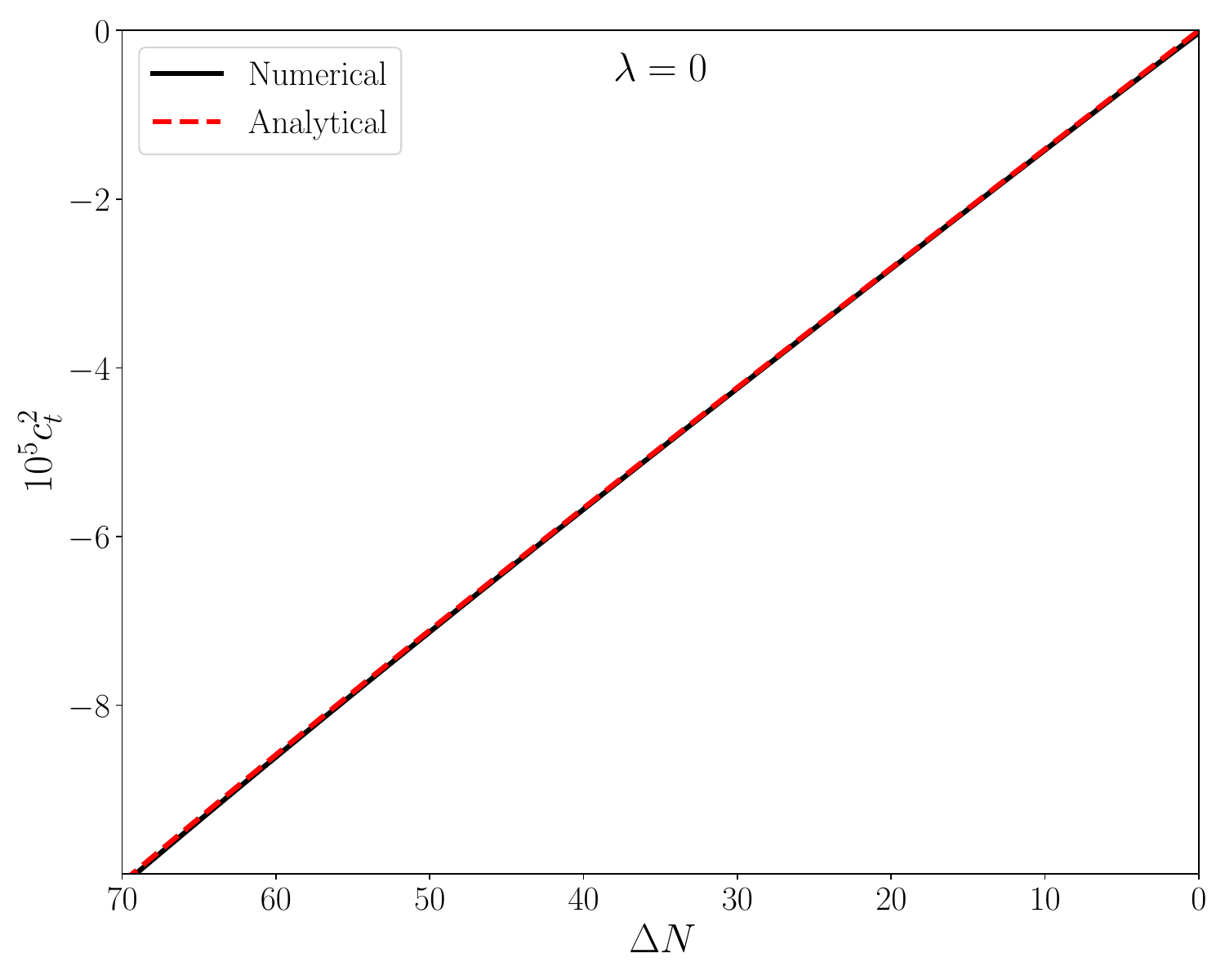}
        \centering
        \label{fig:plot1}
    \end{subfigure}
    \hfill
    \begin{subfigure}[b]{0.45\textwidth}
        \centering
        \includegraphics[scale=0.25]{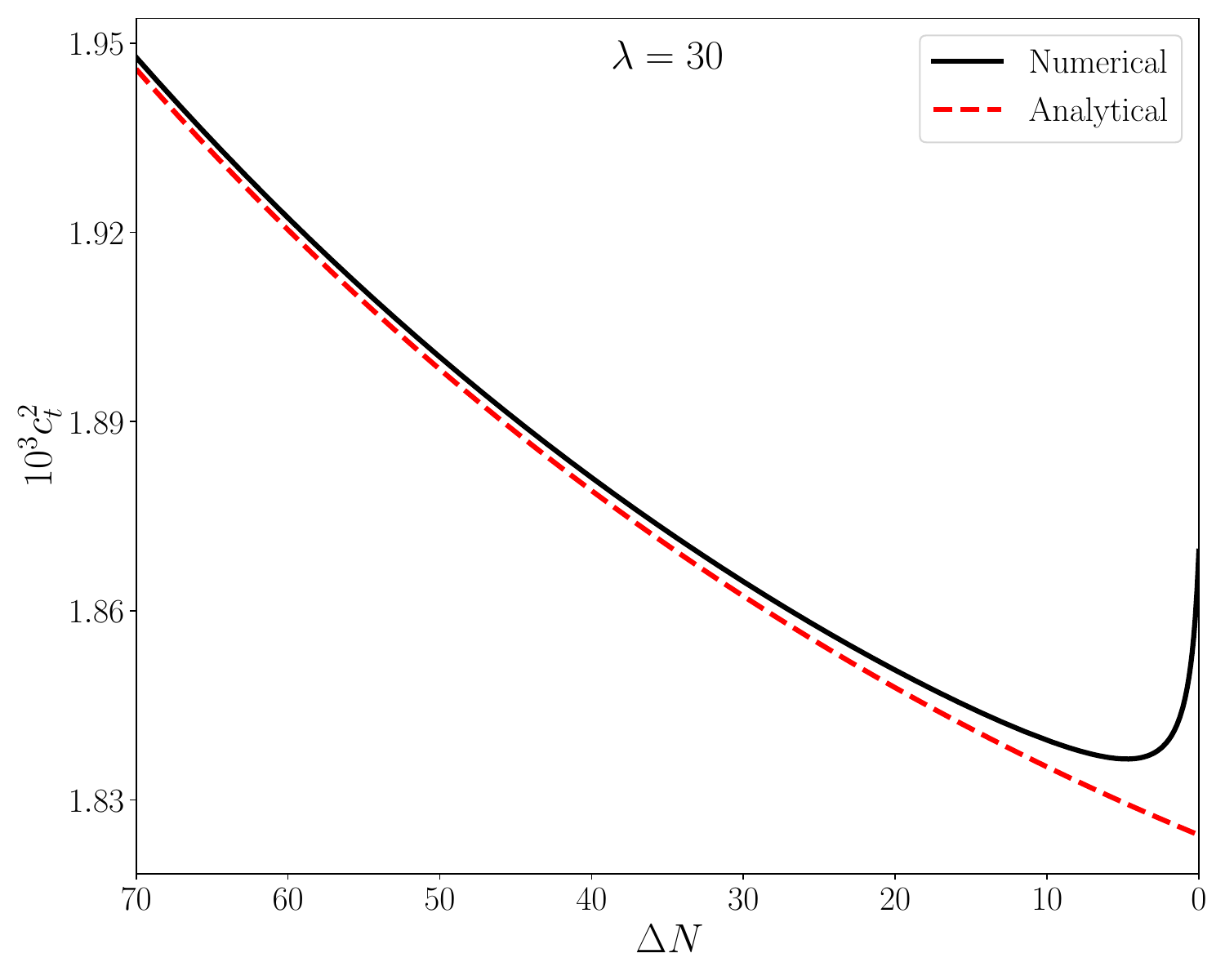}
        \centering
        \label{fig:plot2}
    \end{subfigure}
    \caption{\textit{In the left panel, we show the comparison between the analytical expression for the gauge tensor sound speed Eq.~(\ref{Sound speed analytical}) (dotted red) and the numerical solution (solid black) for $\lambda=0$. The same color scheme applies to the right panel, where we plot the analytical (Eq.~(\ref{analyticalsoundspeedlambda})) vs the numerical result for $\lambda=30$. These plots not only confirm the validity of the analytical approximations but also, in the right panel, display how the introduction of the CS coupling allows for a positive gauge tensor sound speed as long as Eq.~(\ref{ConditionOnLambda}) is satisfied.}}
    \label{Sound speed plots}
\end{figure}

Tensor perturbations for the model  in Eq.~\eqref{GaugeAction} arise from the metric  and gauge fields. We follow here the mode decomposition in \cite{Adshead:2013nka} according to which, without loss of generality, the momentum of each mode is oriented towards the $z$-axis: 
\begin{equation}\label{TensorDecomposition1}
    \delta A_{\mu}^{1}=a(\tau)\left(0,t_{+}(\tau,z),t_{\times}(\tau,z),0\right)\,, \quad \delta A_{\mu}^{2}=a(\tau)\left(0,t_{\times}(\tau,z),-t_{+}(\tau,z),0\right)\,, 
\end{equation}
and
\begin{equation}\label{TensorDecomposition2}
    \delta g_{11}=- \delta g_{22}=\frac{a^{2}(\tau)}{\sqrt{2}}h_{+}(\tau,z)\,, \quad \delta g_{12}=\delta g_{21}=\frac{a^{2}(\tau)}{\sqrt{2}}h_{\times}(\tau,z)\,,
\end{equation}
where  $\tau$ is conformal time. From here on, the prime symbol will represent derivatives w.r.t. to $\tau$. The $+$ and $\times$ symbols stand for the ``plus'' and ``cross'' tensor polarizations. It is convenient to introduce the left ($L$) and right ($R$) polarizations of the gauge field and metric tensor perturbations as linear combinations of the plus and cross polarization states:

\begin{equation}\label{LRtensors}
    h_{L,R}=\frac{h_{+}\pm i h_{\times}}{\sqrt{2}}\,, \quad \quad t_{L,R}=\frac{t_{+}\pm i t_{\times}}{\sqrt{2}}\,.
\end{equation}
Upon plugging Eq.~\eqref{TensorDecomposition1} and Eq.~\eqref{TensorDecomposition2} into Eq.~\eqref{GaugeAction}, going into Fourier space and expanding the action to second order in the perturbations, one arrives at the following action for the canonically-normalized fields:
\begin{equation}\label{TensorAction}
    S_{\Gamma}^{(2)}=S_{L}+S_{R}=\frac{1}{2}\sum_{s=L,R}\int d\tau d^{3}k\left(\Gamma_s^{' \dagger}\Gamma_s'+\Gamma_s^{' \dagger}K_{\Gamma}\Gamma_s-\Gamma_s K_{\Gamma} \Gamma_s^{' \dagger}+\Gamma_s^{\dagger}\Omega_{\Gamma,s}^{2}\Gamma_s\right)\,,
\end{equation} 
where we have used the definitions
\begin{equation}\label{canonicalredefinitiontensors}
    \gamma_s\equiv\begin{pmatrix}
        h_s\\
        t_s\\
    \end{pmatrix}=\mathcal{M}_\Gamma\begin{pmatrix}
        \Psi_s\\
        T_s
    \end{pmatrix}\equiv\mathcal{M}_\Gamma\Gamma_s\,, \quad \mbox{with} \quad \mathcal{M}_\Gamma=\begin{pmatrix}
        \frac{2}{a\Mp} & 0 \\\\
        0 & \frac{1}{\sqrt{2}a}
    \end{pmatrix}\,,
\end{equation}
with $s=L,R$. Here $K_{\Gamma}$ is an antisymmetric matrix, while $\Omega_{\Gamma}$ is symmetric. The associated EoM reads 
\begin{equation}\label{TensorEoM}
    \Gamma_s''+\alpha_{\Gamma}\Gamma_s'+\beta_{\Gamma,s}\,\Gamma_s=0\,, \quad \mbox{with}\quad \alpha_{\Gamma}\equiv 2K_{\Gamma}\,,\quad \beta_{\Gamma,s}\equiv \Omega_{\Gamma,s}^{2}+K_{\Gamma}'\,,
\end{equation}
where we will refer to $\alpha_{\Gamma}$ as derivative interaction matrix and $\beta_{\Gamma}$ as mass matrix. Their explicit form is given by
\begin{equation}\label{alphagamma}
    \alpha_{\Gamma}=\begin{pmatrix}
        0 & \frac{2a\left(HQ+\dot{Q}\right)}{\Mp}\\\\
        -\frac{2a\left(HQ+\dot{Q}\right)}{\Mp} & 0
    \end{pmatrix}\,,
\end{equation}
and
\begin{subequations}
  \begin{align}
    \beta_{\Gamma,11} &= k^{2} + 3a^2\left(\frac{2g^{2} Q^{4}}{\Mp^{2}} + \dot{H}\right) + 2a^2\left(\frac{HQ + \dot{Q}}{\Mp}\right)^{2} + 2a^{2}\left(\frac{\dot{\chi}^2}{\Mp^{2}} - H^{2}\right)\,, \label{betaT11} \\
    \beta_{\Gamma,12} &= \mp \frac{2gkaQ^2}{\Mp}(1-\lambda_1\dot{\chi}^2) - \frac{2g^2a^2Q^3}{\Mp}(1-\lambda_1\dot{\chi}^2)\,, \label{betaT12} \\
    \beta_{\Gamma,21} &= \mp \frac{2gkaQ^2}{\Mp}(1-\lambda_1\dot{\chi}^2) + \frac{2g^2a^2Q^3}{\Mp}(1-\lambda_1\dot{\chi}^2) + \frac{2a^2H}{\Mp}(HQ + \dot{Q})\,, \label{betaT21} \\
    \beta_{\Gamma,22} &= \left(k^{2} \pm 2gkaQ\right)\left(1-\lambda_1\dot{\chi}^2\right)\,, \label{betaT22}  
  \end{align}
\end{subequations} \\
where $\pm$ corresponds to the $L,R$ polarization of the tensor modes, respectively.
From the mass matrix entries in Eqs.~\eqref{betaT11}-\eqref{betaT22} 
one can immediately notice that the gauge field tensor modes have a non-trivial sound speed $c_{t}$ given by
\begin{equation}\label{TensorsoundSpeed}
   c_{t}^{2}\equiv  1-\lambda_{1}\dot{\chi}^{2} \,.
\end{equation}
The approximate solution for $\dot{\chi}$ given in Eq.~(\ref{SlowRollChidot}) then leads to a vanishing sound speed for the gauge field modes. 
On the other hand, relaxing the assumption $H\ll gQ$ in Eq.~\eqref{KGFQEoM} gives the following relation:
\begin{equation}\label{CorrectedQEq}
    2H^{2}Q+2g^{2}Q^{3}\simeq 2g^{2}\lambda_{1}Q^{3}\dot{\chi}^{2}\,,
\end{equation}
which leads to a more accurate result for the sound speed: 
\begin{equation}\label{Sound speed analytical}
    c_t^2\simeq -\frac{H^{2}}{g^{2}Q^{2}}\,,
\end{equation}
as shown in the left panel of Fig. \ref{Sound speed plots}.

\begin{figure}[t]
    \begin{subfigure}[b]{0.45\textwidth}
        \centering
        \includegraphics[scale=0.25]{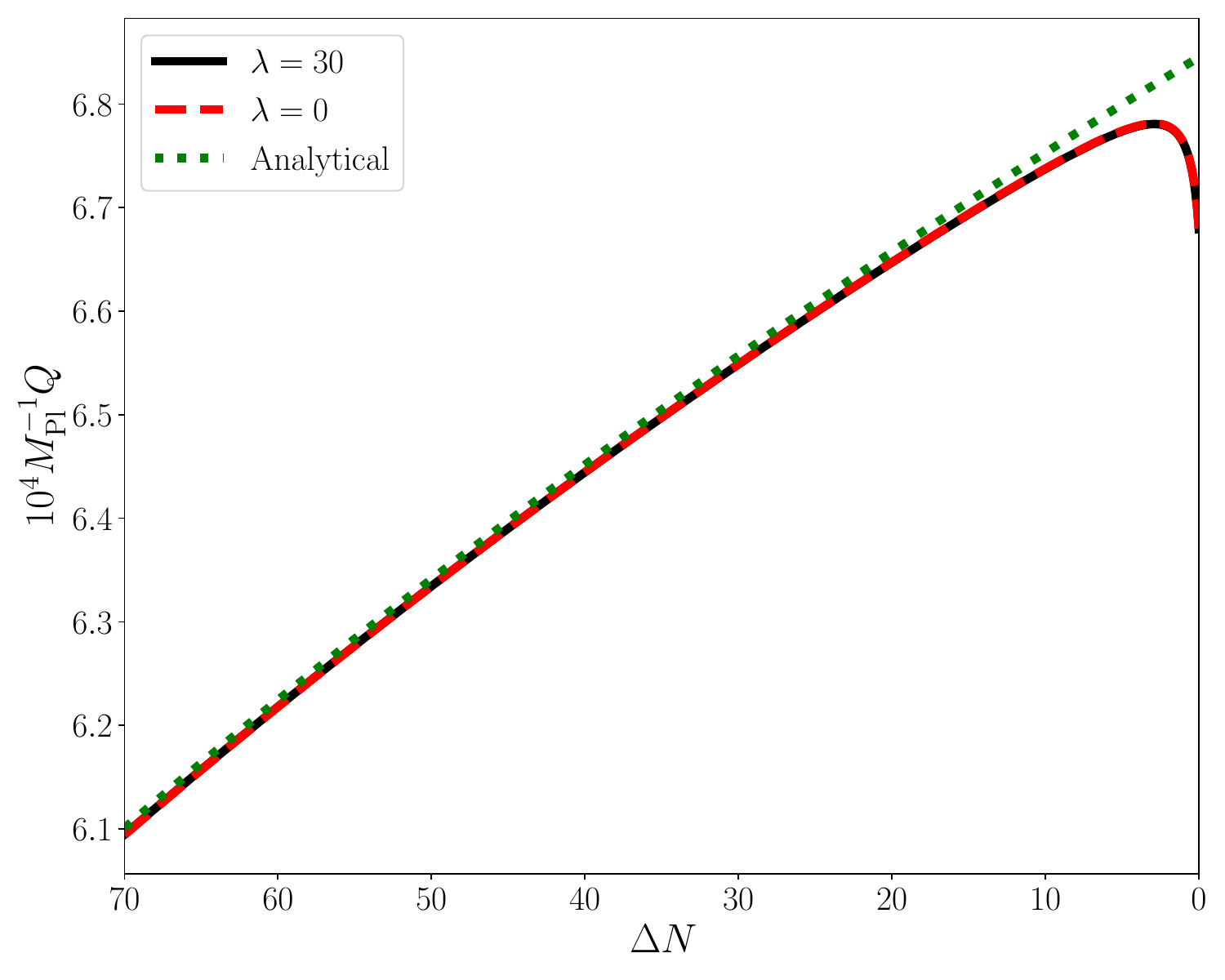}  
        \label{fig:plot2}
    \end{subfigure}
    \hfill
    \begin{subfigure}[b]{0.45\textwidth}
        \centering
        \includegraphics[scale=0.25]{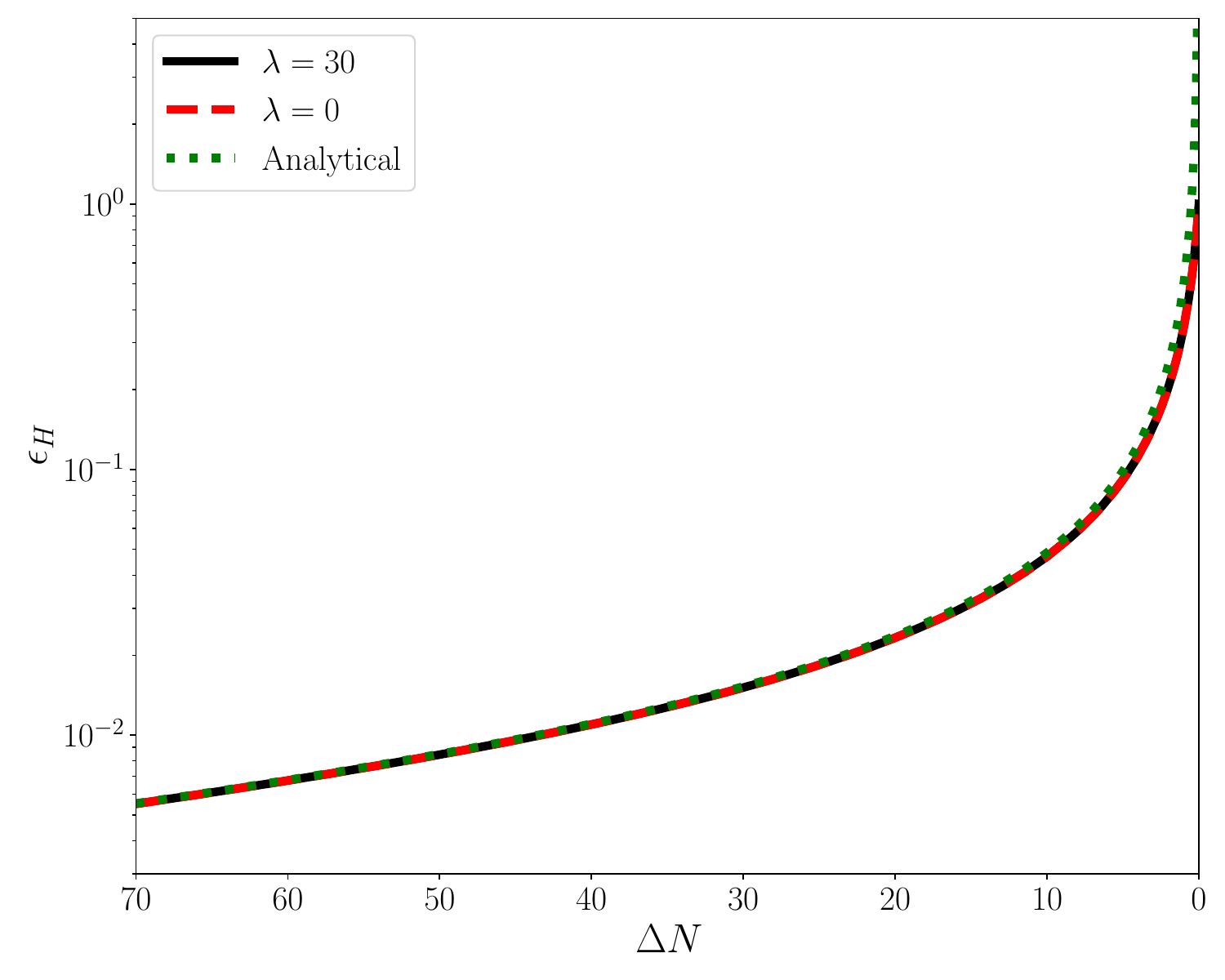} 
        \label{fig:plot4}
    \end{subfigure}

    \caption{\textit{Plots showing the comparison between the numerical evolution of $Q$ and $\epsilon_H$, respectively, for $\lambda=30$ (solid black) and $\lambda=0$ (dashed red), where the remaining parameters are set to the fiducial values in Eq.~(\ref{benchmark}). Additionally, we show in dotted green the approximate solutions in Eq.~(\ref{QN}) and Eq.~(\ref{epsilonN}), showing how these provide a good approximation also for the $\lambda \neq 0$ case and thus implying that the effect of the Chern-Simons coupling remains sub-leading throughout the background evolution. The same conclusion remains valid for the other background quantities, \textit{e.g.} $\chi$ and $H$.}}
    \label{fig:comparison_lambda}
\end{figure}

\indent A negative sound speed for the gauge tensor modes urges us to modify our original proposal, Eq.~\eqref{GaugeAction}. We thus relax the parity-even requirement in order to consider the addition of the Chern-Simons operator to Eq.~\eqref{GaugeAction}\footnote{The Chern-Simons term is the lowest-dimension operator compatible with the shift symmetry.}: 
\begin{equation}\label{GaugeAction2}
\begin{split}
    S=\int d^{4}x\sqrt{-g}&\Bigg[\frac{\Mp^2}{2}R-\frac{1}{2}\left(g^{\mu\nu}+\lambda_{1}\left(F^{a\mu}_{\ \ \ \alpha}F^{a\alpha\nu}+\frac{1}{2}F^{a\rho\sigma}F^{a}_{\ \rho\sigma}g^{\mu\nu}\right)\right)\partial_{\mu}\chi\partial_{\nu}\chi\\
    &-V(\chi)-\frac{1}{4}F^{a}_{\mu\nu}F_{a}^{\mu\nu}- \frac{\lambda}{8f\sqrt{-g}}\chi\epsilon^{\mu\nu\sigma\rho} F^{a}_{\mu\nu} F^{a}_{\rho\sigma}\Bigg]\, ,
\end{split}
\end{equation}
ending up with what we call  \textit{kinetic gauge friction chromo-natural} (KFC) inflation.
As we will prove shortly, the added term allows for a positive sound speed. Before studying the fluctuations for the new Lagrangian, let us briefly comment on the background evolution. The new background EoM read
\begin{equation}\label{KGFXEoM2}
    \ddot{\chi}\left(1+3g^{2}\lambda_{1}Q^{4}\right)+3H\dot{\chi} \left(1+\frac{4g^{2}\lambda_{1}Q^{3} \dot{Q}}{H}+3g^{2} \lambda_{1}Q^4\right)+V_{\chi}(\chi)=-\frac{3g\lambda Q^2}{f}\left(\dot{Q}+HQ\right)\,,
\end{equation}
\begin{equation}\label{KGFQEoM2}
    \ddot{Q}+3H\dot{Q}+Q\left(\dot{H}+2 H^{2}\right)+2g^{2}Q^{3}=2g^{2}\lambda_{1} Q^{3}\dot{\chi}^{2}+\frac{g\lambda}{f}\dot{\chi}Q^{2}\,.
\end{equation}
We verified that, for $\lambda\ll fg\lambda_1Q\dot{\chi}$, the background evolution of the full model is well approximated by the one previously obtained in the $\lambda=0$ case (see Fig.~\ref{fig:comparison_lambda}). 
The results obtained in Sec.~\ref{Section background} are therefore still valid in this regime. \\
\indent Moving on to perturbations, the entries of the mass matrix in Eq.~\eqref{TensorEoM} are now modified as
\begin{equation}\label{betaGammaLambda}
    \begin{split}
        &\beta_{\Gamma, 11}\xrightarrow[]{}\beta_{\Gamma, 11}\, ,\\
        &\beta_{\Gamma, 12}\xrightarrow[]{}\beta_{\Gamma, 12}\, ,\\
        &\beta_{\Gamma, 21}\xrightarrow[]{}\beta_{\Gamma, 21}-\frac{2g\lambda a^2Q^2\dot{\chi}}{\Mp f}\, ,\\
        &\beta_{\Gamma, 22}\xrightarrow[]{}\beta_{\Gamma, 22}\pm\frac{\lambda ka\dot{\chi}}{f}+\frac{g\lambda a^2Q\dot{\chi}}{f}\, ,
    \end{split}
\end{equation}
while the $\alpha_{\Gamma}$ is unaffected. Notably, the inclusion of the CS coupling does not explicitly change the expression for the
sound speed of the  gauge tensors in Eq.~\eqref{TensorsoundSpeed}. Although the formula for the sound speed does not explicitly depend on $\lambda$, the dynamics of the axion (and gauge field) background do. As we shall show next, this dependence will allow for a positive sound speed within certain ranges of the model parameters.
From $\eqref{KGFQEoM2}$, to leading order in slow-roll one finds
\begin{equation}
    2H^2Q+2g^2Q^3\simeq\frac{g\lambda}{f}\dot{\chi}Q^2+2g^2\lambda_1Q^3\dot{\chi}^2\, ,
\end{equation}
from which one can solve in terms of $\dot{\chi}$ and replace it in the expression for the sound speed, obtaining
\begin{equation}\label{analyticalsoundspeedlambda}
    c_{t}^{2}\simeq-\frac{H^{2}}{g^{2}Q^{2}}+\frac{\lambda}{8f^{2}g^{2}\lambda_{1}Q^{2}}\left(\sqrt{\lambda^{2}+16f^{2}\lambda_{1}\left(H^{2}+g^{2}Q^{2}\right)}-\lambda\right)\, .
\end{equation}
Requiring a positive sound speed then gives a lower bound on the CS coupling constant $\lambda$:
\begin{equation}\label{ConditionOnLambda}
    \lambda\gtrsim\frac{2 f \sqrt{\lambda_{1}} H^2}{gQ}\, .    
\end{equation}
In the right panel of Fig.~$\ref{Sound speed plots}$ we show the evolution of the analytical expression for the sound speed in Eq.~\eqref{analyticalsoundspeedlambda}, together with its numerical evaluation, for our choice of the fiducial parameters, now including also a value for $\lambda$\footnote{The chosen $\lambda$ sits in the range $\frac{2 f \sqrt{\lambda_{1}} H^2}{gQ}<\lambda\ll fg\lambda_1Q\dot{\chi}$ guaranteeing both a positive sound speed for the gauge field tensors and a negligible background contribution from the CS term. Note that a small value of $\lambda$ is sufficient to ensure that the overall value of $c_t$ stays positive.}:
\begin{equation}
\label{benchmark2}
\{ g= 0.5, \,\lambda_{1}= \left(1\times 10^{4}\Mp^{-1}\right)^{4},\, \mu=2.2\times10^{-3}\Mp, \,f= 0.24\Mp,\,\lambda=30 \}\,.
\end{equation}
We shall then use these parameters in the remainder of the text.

\subsection{Quantization and initial conditions}

The next step is to determine the initial conditions for $\Gamma_{s}$. We begin by expanding into creation and annihilation operators 
\begin{equation}
    \Gamma_{i}(\tau,\mathbf{k})=\mathcal{G}_{ij}(\tau,k)a_{j}(\mathbf{k})+\mathcal{G}_{ij}^{*}(\tau,k)a^{\dagger}_{j}(-\mathbf{k})\,,
\end{equation}
where the mode functions $\mathcal{G}_{ij}$ satisfy the classical EoM:
\begin{equation} \label{TensorModesEoM}
    \mathcal{G}_{ij}''+\alpha_{\Gamma}\mathcal{G}_{ij}'+\beta_{\Gamma}\mathcal{G}_{ij}=0\,,
\end{equation}
where the polarization index $s$ is kept implicit. This can be written explicitly in terms of the canonically-normalized gauge field and metric polarizations as
\begin{equation}\label{PsiEq}
\begin{split}
    \partial_{x}^{2}\Psi_{L,R}&+\left(1-\frac{2}{x^2}-\frac{3\epsilon_H}{x^2}+\frac{6\epsilon_{Q}}{x^{2}}+\frac{2\epsilon_{Q}}{m_{Q}^{2}x^{2}}+\frac{8\epsilon_{Q}\xi^{2}}{m_{Q}^{2}\Lambda^{2}x^{2}}\right)\Psi_{L,R} \\
    &=\frac{2\sqrt{\epsilon_{Q}}}{m_{Q}x}\partial_{x}T_{L,R}+\frac{2\sqrt{\epsilon_{Q}}}{x^{2}}(m_Q\pm x)\left(1-\frac{4G\xi^2}{m_Q^2\Lambda^2}\right)T_{L,R}\,,
\end{split}
\end{equation}
and
\begin{equation}\label{TEq}
\begin{split}
    \partial_{x}^{2}T_{L,R}&+\left[\left(1\pm\frac{2m_{Q}}{x}\right)\left(1-\frac{4G\xi^2}{m_Q^2\Lambda^2}\right)+\frac{2\xi}{x^2}(m_Q\pm x)\right]T_{L,R} \\
    &=-\frac{2\sqrt{\epsilon_{Q}}}{m_{Q} x}\partial_{x}\Psi_{L,R}-\frac{2\sqrt{\epsilon_{Q}}}{x^2}\left((m_Q\mp x)\left(1-\frac{4G\xi^2}{m_Q^2\Lambda^2}\right)+\frac{1}{m_Q}(1-2m_Q\xi)\right)\Psi_{L,R}\,,
\end{split}
\end{equation}
where we moved to the dimensionless time variable $x\equiv -k\tau$. In these equations, we used the fact that $\dot{Q}\ll HQ$ and introduced the following dimensionless quantities:
\begin{equation}\label{par}
    m_Q
\equiv \frac{gQ}{H}\,, \quad \xi\equiv\frac{\lambda\dot{\chi}}{2fH}\,, \quad  G\equiv g^2\lambda_1Q^4\,, \quad \Lambda\equiv \frac{\lambda Q}{f}\,,\quad  \epsilon_Q\equiv \frac{g^2Q^4}{H^2}\,.
\end{equation} 
One can notice that the EoM in Eqs~\eqref{PsiEq}-\eqref{TEq} consistently reduce to the ones in chromo-natural inflation once the terms originating from the kinetic coupling are switched off (\textit{i.e.} $G\xrightarrow[]{}0$).\\\indent In the infinite past limit $(x\xrightarrow[]{}\infty)$ the EoM read
\begin{equation}\label{TensorEoMearlyt}
    \partial_{x}^{2}\Psi_{L,R}+\Psi_{L,R}=0\,, \qquad \partial_{x}^{2}T_{L,R}+T_{L,R}\left(1-\frac{4G\xi^2}{m_Q^2\Lambda^2}\right)=0\, .
\end{equation}
whose solutions correspond to the Bunch-Davies vacua, with a non-trivial sound speed in the case of the gauge field perturbations\footnote{The sound speed $c_t$ was introduced in Eq.~(\ref{TensorsoundSpeed}) for the first time and it is here rewritten in terms of the dimensionless background quantities given in Eq.~(\ref{par}).}:
\begin{equation}\label{tensorSoundSpeedG}
    c_t^2=1-\frac{4G\xi^2}{m_Q^2\Lambda^2}\,.
\end{equation}
Consequently, the initial conditions for the tensor modes  are given by
\begin{equation}\label{initialtensors}
   \sqrt{2k} \mathcal{G}_{\rm in}=\begin{pmatrix}
        e^{ix} & \quad 0\\
        0 & \quad c_t^{-1/2}e^{ic_tx}
    \end{pmatrix}_{x=x_{\rm in}}\,,  \quad \sqrt{2k}(\partial_{x}\mathcal{G})_{\rm in}=i\begin{pmatrix}
        e^{ix} & \quad 0\\
        0 & \quad c_t^{1/2}e^{ic_tx}
    \end{pmatrix}_{x=x_{\rm in}}\,.
\end{equation}
\indent Once the solution for the mode functions of the canonically-normalized fields is obtained, we can rotate back to the physical tensor perturbations by using Eq.~\eqref{canonicalredefinitiontensors}. The corresponding dimensionless power spectra can then be written as
\begin{equation}\label{tensorspectra}
    \mathcal{P}^{(T)}_{ij}=\frac{k^3}{2\pi^2}\sum_{s=L,R}{\rm Re}\left[\mathcal{M}_{\Gamma,ik}\mathcal{M}_{\Gamma,j\ell}\mathcal{G}_{km}\mathcal{G}^*_{\ell m}\right]\,.
\end{equation}
We will make use of this formula to compute the observables.
\subsection{Analytical solution for the enhanced gauge tensor polarization}
\begin{figure}[t!]
    \includegraphics[scale=0.35]{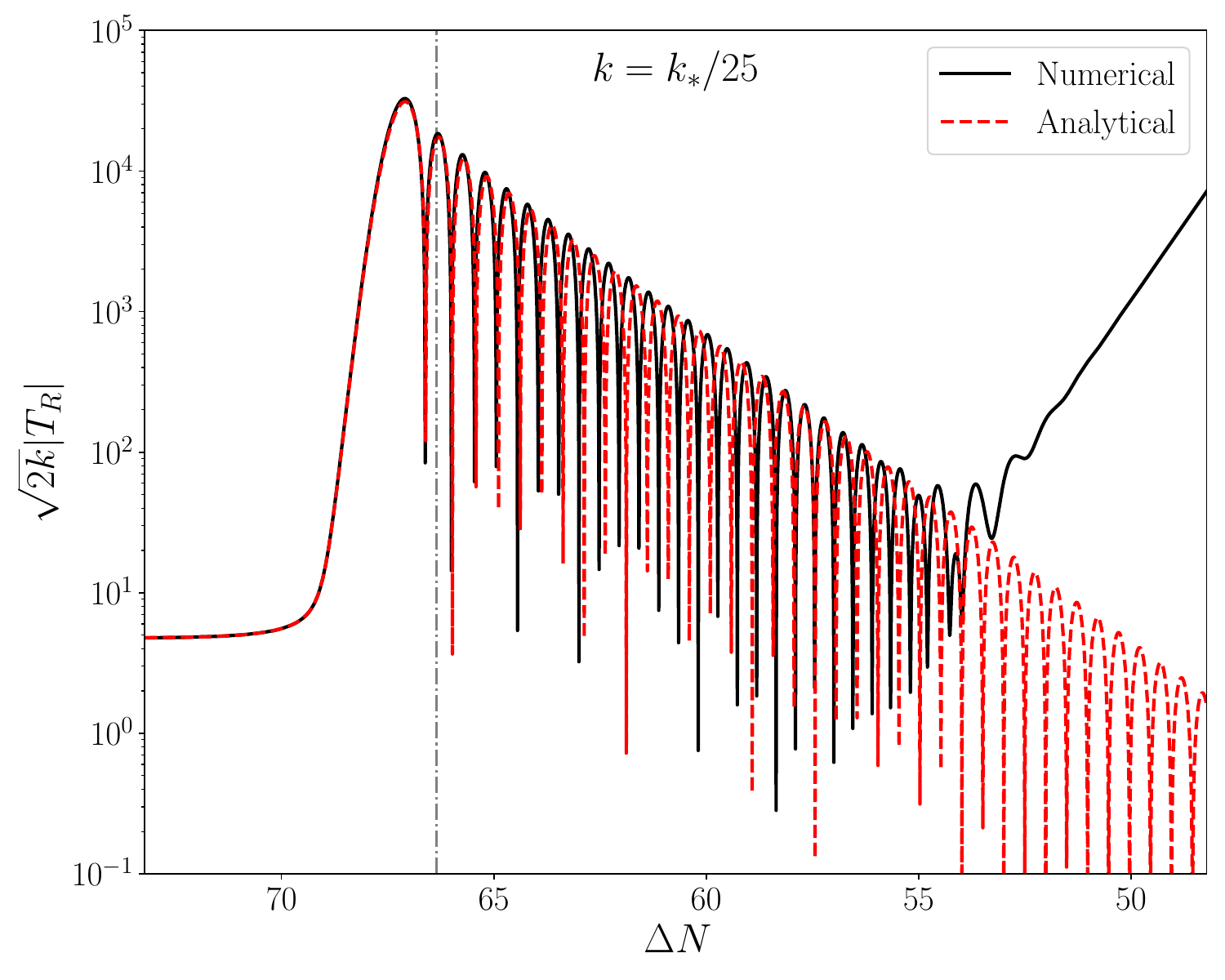}
    \caption{\textit{Comparison between numerical (solid black) and analytical (dashed red) solution of the enhanced gauge tensor polarization $T_R$ for the fiducial set of parameters in Eq.~(\ref{benchmark2}). As expected, the change in behaviour of the mode due to the mixing with metric tensors on super-horizon scales is not accounted for by the analytical result. However, the latter is in good agreement with the numerical result  around the time of maximum enhancement of the mode before horizon-crossing (vertical dash-dotted line).}}\label{analytical tensor}
\end{figure} 

Following the treatment in CNI \cite{Adshead:2013nka}, it is possible to analytically solve Eq.~\eqref{TEq} for the enhanced polarization $R$ by neglecting the mixing terms with the metric tensors:
\begin{equation}\label{TEqAn}
    \partial_{x}^{2}T_{R}+\left[\left(1-\frac{2m_{Q}}{x}\right)c_t^2+\frac{2\xi}{x^2}(m_Q- x)\right]T_{R}\simeq0\,,
\end{equation}
where the background quantities $m_Q$, $\xi$ and $c_t$ are assumed to be slowly-varying. In particular, their value is set to the one at horizon-crossing, \textit{i.e.} at $c_tx=1$. The solution can then be written as
\begin{equation}\label{TRanalytical}
    T_R(x)= A_k M_{\alpha,\beta}(2ic_tx)+B_k W_{\alpha,\beta}(2ic_tx)\,.
\end{equation}
Here $M, W$ are Whittaker functions and 
\begin{equation}
    \alpha\equiv i(c_t^{-1}\xi+c_tm_Q), \quad \beta\equiv\sqrt{\frac{1}{4}-2\xi m_Q}\,,
\end{equation}
where $c_{t}$ is given in Eq.~\eqref{tensorSoundSpeedG}. The integration constants $A_k,B_k$ are then fixed by requiring that, in the infinite past, the solution matches the initial conditions given in Eq.~\eqref{initialtensors}, leading to
\begin{equation}
\begin{split}
    A_k=\frac{1}{\sqrt{2c_tk}}\frac{\Gamma \left(-\alpha+\beta+\frac{1}{2}\right)}{(2 i)^{-\alpha} \Gamma (2 \beta+1)}\,, \quad
     B_k=-\frac{1}{\sqrt{2c_tk}}\frac{\Gamma \left(-\alpha+\beta+\frac{1}{2}\right)}{(2 i)^{-\alpha} \Gamma (\alpha+\beta+\frac{1}{2})}2^\alpha i^{\beta+1}(-i)^{\alpha-\beta}\,. \\
\end{split}
\end{equation}
We plot this analytical solution in Fig.~\ref{analytical tensor}, alongside the numerical one obtained by solving the coupled system of Eqs.~(\ref{PsiEq})-(\ref{TEq}). 

Eq.~\eqref{TRanalytical} can also be written as $\sqrt{2c_tk}T_R=(-2i)^\alpha W_{-\alpha,\beta}(-2ic_tx)$, from which one can verify that the amplitude of the enhanced gauge mode grows in the following way:
\begin{equation}\label{TRamplitude}
    |T_R|\propto c_t^{-1/2}\exp\left[\frac{\pi}{2}(c_t^{-1}\xi+c_tm_Q)\right]\,,
\end{equation}
so that reducing $c_t$ makes $|T_R|$ larger/smaller when $\xi\gtrless c_t^2 m_Q-c_t/\pi$, respectively.
\subsection{Numerical evolution}
\begin{figure}[t!]
    \includegraphics[scale=0.35]{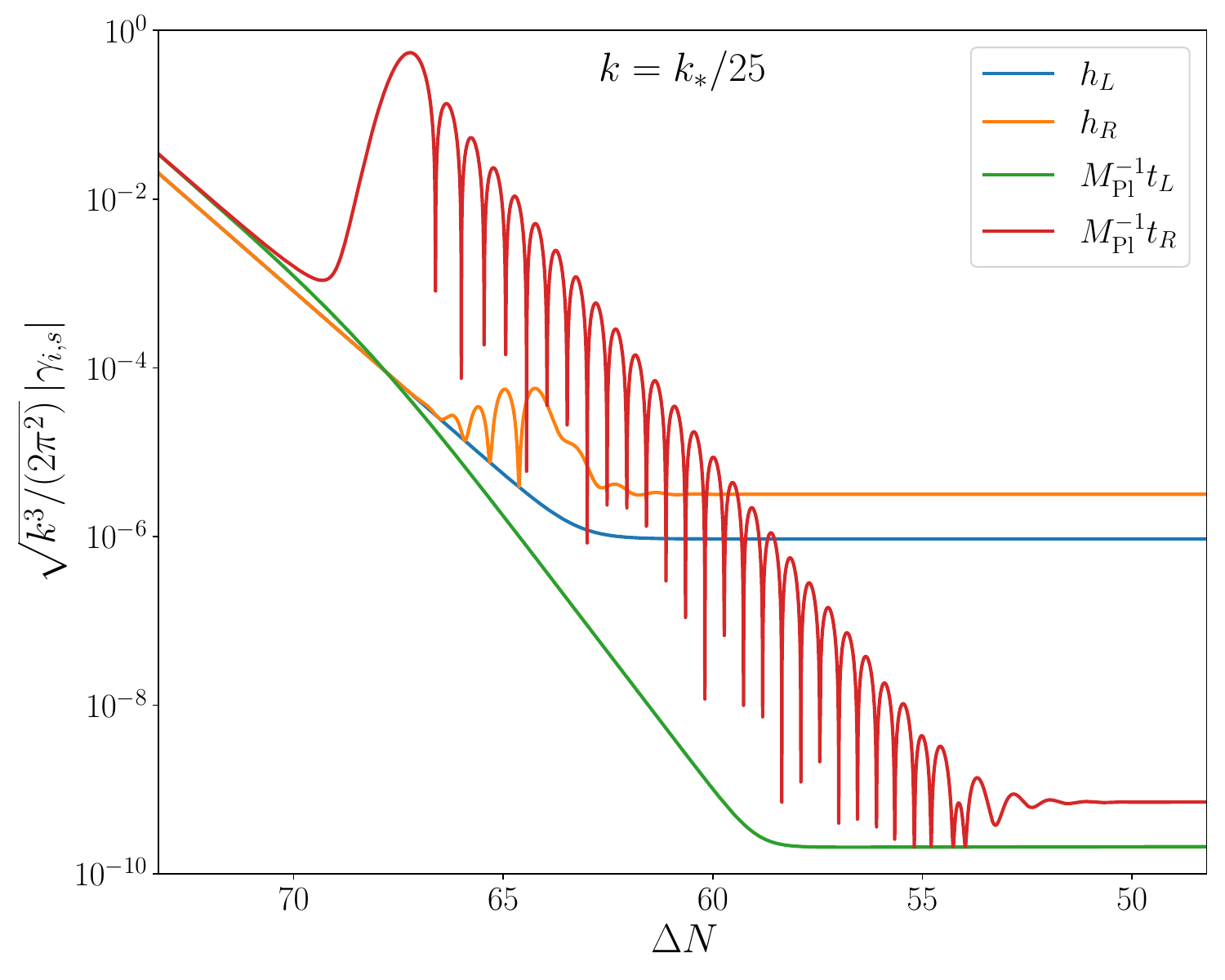}
    \caption{\textit{Left- and right-handed  mode functions for the tensors modes, obtained by numerically solving the system in Eq.~(\ref{TensorEoM}), considering the initial conditions Eq.~(\ref{initialtensors}) together with the fiducial set of parameters from Eq.~(\ref{benchmark2}). Here we are showing the $k$-modes with $k=k_*/25$, where $k_*$ crosses the horizon at the CMB scale, conventionally set to be at sixty $e$-folds before the end of inflation. From the plot it is clear that a chiral source of GWs is present, in analogy with CNI and related models.}} \label{numerical tensor}
\end{figure}

The power spectrum for the metric tensor modes   ($h$) is comprised by both a vacuum contribution and a sourced one
\begin{equation}
    \mathcal{P}_{h}^{\rm tot}=\mathcal{P}_{h}^{\rm vacuum}+\mathcal{P}_{h}^{\rm sourced}\,.
\end{equation}
From Eq.~\eqref{tensorspectra}, one finds
\begin{equation}
   \mathcal{P}_{h}^{\rm tot}=\frac{k^{3}}{2\pi^{2}}\left(\frac{2}{a\Mp}\right)^{2}\Big(|\Psi_{L}|^{2}+|\Psi_{R}|^{2}\Big)\, .
\end{equation} 
It is important to stress at this stage that, whenever the sourced contribution to the GW spectrum is the leading one (as in our case), the signal will be chiral, a feature that can be tested both at large and intermediate/small scales \cite{Seto:2007tn,Smith:2016jqs,Thorne:2017jft,Domcke:2019zls}. In Fig.~\ref{numerical tensor} we plot the numerical solutions of the individual polarizations as functions of the number of $e$-folds.
\section{Scalar perturbations}\label{Kinetic gauge friction: Scalar perturbations}

We now consider the perturbations in the scalar sector of the model, restricting to the ones in the axion and gauge fields, while neglecting scalar perturbations in the metric.  This is a good approximation that has been shown to be valid  for several existing models in axion inflation (see \textit{e.g.} \cite{Dimastrogiovanni:2012ew}). Let us adopt the decomposition of the gauge field from \cite{Papageorgiou:2018rfx}, given by  
\begin{equation}\label{ScalarDecomposition}
    \delta A_{\mu}^{1}=\left(0,\delta\phi(\tau,z)-\delta Z(\tau,z),\chi_{3}(\tau,z),0\right)\,,
\end{equation}
\begin{equation*}
    \delta A_{\mu}^{2}=\left(0,-\chi_{3}(\tau,z),\delta\phi(\tau,z)-\delta Z(\tau,z),0\right)\,,
\end{equation*}
\begin{equation*}
    \delta A_{\mu}^{3}=\left(\delta A_{0}^{3}(\tau,z),0,0,\delta\phi(\tau,z)+2\delta Z(\tau,z)\right)\,.
\end{equation*}
Plugging this into Eq.~\eqref{GaugeAction2}, going into Fourier space and expanding to second order in the perturbations, one finds that out of the five modes (four in the gauge sector and $\delta\chi$ from the axion) one is non-dynamical, namely $\delta A_{0}^{3}$. Furthermore, we impose the following gauge condition \cite{Papageorgiou:2018rfx}:
\begin{equation}\label{gaugechoice}
    \chi_{3}=-ik\left(\frac{2\delta Z+\delta\phi}{2gaQ}\right)\,,
\end{equation} 
which can then be used when solving the algebraic constraint equation for $\delta A_{0}^{3}$ in terms of the dynamical modes, obtaining 
\begin{equation}\label{A0}
    \delta A_{0}^{3}=-ik\left(\frac{gaQ^{3}\delta\chi(\lambda a+2f g\lambda_{1}Q\chi')-2f(2\delta Z+\delta\phi)(Q'+aQH)}{fQ(k^{2}+2g^{2}a^{2}Q^{2})}\right)\,.
\end{equation}
With Eqs.~\eqref{gaugechoice} and \eqref{A0}, the action at second order in the perturbations can be fully expressed in terms of the dynamical modes $\delta\chi$, $\delta\phi$ and $\delta Z$, which can be related to the canonically-normalized fields as

\begin{equation}\label{canonicalredefinitionscalars}
    \delta\equiv\begin{pmatrix}
        &\delta\chi\\
        &\delta\phi\\
        &\delta Z
    \end{pmatrix}=\mathcal{M}_\Delta\begin{pmatrix}
        &X\\
        &\varphi\\
        &Z
    \end{pmatrix}\equiv\mathcal{M}_\Delta\Delta\,,
\end{equation}
with
\begin{equation}\label{kineticmatrix}
\mathcal{M}_\Delta=\begin{pmatrix}
    \frac{1}{a\sqrt{1+3g^{2}\lambda_{1}Q^{4}}} & 0\quad & 0 \\\\
    0 & \quad \frac{\sqrt{2}g a Q}{3 \sqrt{k^2+2g^2 a^2 Q^2}} & \quad \frac{\sqrt{2}}{3} \\\\
    0 & \frac{\sqrt{2}g a Q}{3 \sqrt{k^2+2g^2 a^2 Q^2}} & \quad -\frac{1}{3\sqrt{2}}
\end{pmatrix}\,.
\end{equation}
The action for the canonically-normalized scalar modes $\Delta$ then reads
\begin{equation}\label{ScalarAction}
    S_{\Delta}^{(2)}=\frac{1}{2}\int d\tau d^{3}k\left(\Delta^{' \dagger}\Delta'+\Delta^{' \dagger}K_{\Delta}\Delta-\Delta K_{\Delta} \Delta^{' \dagger}+\Delta^{\dagger}\Omega_{\Delta}^{2}\Delta\right)\,.
\end{equation}
As in the tensor case, $K_{\Delta}$ and $\Omega_{\Delta}$ are antisymmetric and symmetric matrices, respectively. The corresponding EoM are given by:
\begin{equation}\label{ScalarEoM}
    \Delta''+\alpha_{\Delta}\Delta'+\beta_{\Delta}\Delta=0\,, \quad \mbox{with} \quad \alpha_{\Delta}\equiv 2K_{\Delta}\,, \quad \beta_{\Delta}\equiv \Omega_{\Delta}^{2}+K_{\Delta}'\,. 
\end{equation}
where the complete expressions of $\alpha_{\Delta}$ and $\beta_{\Delta}$ can be found in Appendix~\ref{app:B}.
Here, in the main text, we only report the infinite past $(x\xrightarrow[]{}\infty)$ limit of these matrices, needed in order to set the initial conditions for the scalar modes:
\begin{equation}\label{alphadelta}
    \alpha_{\Delta, \rm in}=\begin{pmatrix}
         0 &\quad -\frac{2\sqrt{2}G\xi}{m_{Q}\Lambda\sqrt{1+3G}} & \quad 0\\\\
         \frac{2\sqrt{2}G\xi}{m_{Q}\Lambda\sqrt{1+3G}} &0\quad &  \quad 0\\\\
        0 &0\quad  & \quad 0\\
    \end{pmatrix}\,, 
\end{equation}
and
\begin{equation}\label{betadelta}
    \beta_{\Delta, \rm in}=
    \begin{pmatrix}
      \frac{m_Q^2+G(m_Q^2-2)}{(1+3G)m_Q^2} & 0 & -\frac{2\sqrt{2}G\xi}{\sqrt{1+3G}m_Q^2\Lambda} \\\\
    0 & 1-\frac{4G\xi^2}{m_Q^2\Lambda^2} & 0 \\\\
-\frac{2\sqrt{2}G\xi}{\sqrt{1+3G}m_Q^2\Lambda} & 0 & 1-\frac{4G\xi^2}{m_Q^2\Lambda^2} 
    \end{pmatrix}\,.
\end{equation}
Given that Eq.~\eqref{alphadelta} has non-zero elements, and Eq.~\eqref{betadelta} is non-diagonal, it is clear that the system does not decouple in the infinite past, leading to an early time solution different from the standard Bunch-Davies one.\\

\subsection{Multifield quantization and the Wronskian condition}\label{Initial conditions}
The triplet $\Delta$ is quantized in terms of mode functions $\mathcal{D}_{ij}$ as
\begin{equation}\label{DeltaDecom}
    \Delta_{i}(\tau,\mathbf{k})=\mathcal{D}_{ij}\left(\tau,k\right)a_{j}(\mathbf{k})+\mathcal{D}^{*}_{ij}\left(\tau,k\right)a^{\dagger}_{j}(\mathbf{-k})\,,
\end{equation}
where creation and annihilation operators follow the usual commutation relation, $\left[a_{i}\left(\textbf{k}\right),a_{j}^{\dagger}\left(\textbf{k}'\right)\right]=\delta^{(3)}\left(\textbf{k}-\textbf{k}'\right)\delta_{ij}$, and the mode functions obey the classical EoM:
\begin{equation} \label{ScalarModesEoM}
    \mathcal{D}_{ij}''+\alpha_{\Delta}\mathcal{D}_{ij}'+\beta_{\Delta}\mathcal{D}_{ij}=0\,.
\end{equation}
One also needs to impose the equal-time commutation relations between the triplet $\Delta_{i}$ and its conjugate momentum $\Pi_{i}$: 
\begin{equation}\label{equaltimecommu}
    \left[\Delta_{i}(\tau,\textbf{x}),\Pi_{j}(\tau,\textbf{x}')\right]=i\delta^{(3)}(\textbf{x}-\textbf{x}')\delta_{ij}\,, \quad \mbox{with} \quad \Pi_{i}\equiv\frac{\partial L}{\partial \Delta'_{i}}\,,
\end{equation}
where, similarly to $\Delta_{i}$, the conjugate momentum $\Pi_i$ is expanded as 
\begin{equation}
    \Pi_{i}\equiv\pi_{ij}a_{j}+\pi^{*}_{ij}a^{\dagger}_{j}\,, \quad \mbox{with}  \quad\pi_{ij}=\mathcal{D}'_{ij}+\frac{\alpha_{\Delta}}{2}\mathcal{D}_{ij}\,.
\end{equation}
It is then straightforward to show that the commutation relation for the creation/annihilation operators and Eq.~\eqref{equaltimecommu} are simultaneously satisfied by requiring
\begin{equation}\label{MultifieldWronsk}
    \left[\mathcal{D}\pi^{\dagger}-\mathcal{D}^{*}\pi^{T}\right]_{ij}=i\delta_{ij}\,,
\end{equation}
which is the multifield version of the Wronskian condition. However, unlike the single-field case where the EoM ensures that Eq.~\eqref{MultifieldWronsk} is satisfied at all times, in the present case one also needs to impose the relations \begin{equation}\label{MultifieldWronsk2}
    \pi\pi^{\dagger}-\pi^{*}\pi^{T}=\mathcal{D}\mathcal{D}^{\dagger}-\mathcal{D}^{*}\mathcal{D}^{T}=0\,,
\end{equation}
that are satisfied if the products $\mathcal{D}\mathcal{D}^{\dagger}$ and $\pi\pi^{\dagger}$ are real. We will refer to the set of Eq.~\eqref{MultifieldWronsk} and Eq.~\eqref{MultifieldWronsk2} as the generalized Wronskian conditions \cite{Namba:2013kia}. \\
\indent The quantization procedure can  be then summarized in two steps: (i) classically solve the system in the infinite past limit and (ii) impose the generalized Wronskian condition to determine all integration constants. 
\subsection{Solution of the infinite past system}
We will now derive the initial conditions for scalar perturbations by solving the system of equations, given by Eq.~\eqref{ScalarEoM}, in the infinite past limit. We first rewrite it explicitly in terms of time variable $x$:
\begin{equation}\label{infinitePastScalarEoM}
\begin{aligned}
    &\partial_{x}^{2}X_{\rm in}+\left(\frac{m_Q^2+G(m_Q^2-2)}{(1+3G)m_Q^2}\right)X_{\rm in}-\frac{2\sqrt{2}G\xi}{m_Q^2\Lambda\sqrt{1+3G}}Z_{\rm in}-\frac{2\sqrt{2}G\xi}{m_{Q}\Lambda\sqrt{1+3G}}\partial_{x}\varphi_{\rm in}=0\,, \\
    &\partial_{x}^{2}\varphi_{\rm in}+ \left(1-\frac{4G\xi^2}{m_Q^2\Lambda^2}\right)\varphi_{\rm in}+ \frac{2\sqrt{2}G\xi}{m_{Q}\Lambda\sqrt{1+3G}}\partial_{x}X_{\rm in}=0\,, \\
    &\partial_{x}^{2}Z_{\rm in}+\left(1-\frac{4G\xi^2}{m_Q^2\Lambda^2} \right)Z_{\rm in}-\frac{2\sqrt{2}G\xi}{m_Q^2\Lambda\sqrt{1+3G}}X_{\rm in}=0\,.
\end{aligned}
\end{equation}
In order to derive an analytical solution, it is convenient to rewrite the $3$D second-order system as a $6$D  first-order one by means of the following change of variables:
\begin{equation}
    \partial_{x}X_{\rm in}=\rho\,, \quad \partial_{x}\varphi_{\rm in}=\sigma\,,\quad \partial_{x}Z_{\rm in}=\gamma\,.
\end{equation}
The 6D system, in matrix form, is then given by
\begin{equation}\label{higherdimsystem}
    \mathcal{R}'_{\rm in}=\mathcal{N}\mathcal{R}_{\rm in}\,, \quad \mbox{where}\quad\mathcal{R}_{\rm in}=\begin{pmatrix}
        X_{\rm in },&\rho,&\varphi_{\rm in },&\sigma,&Z_{\rm in },&\gamma
    \end{pmatrix}^{T}\,,
\end{equation}
with
\begin{equation}\label{Coefficienteq}
\mathcal{N}=\begin{pmatrix}
        0&1&0&0&0&0\\
        
        -\frac{m_Q^2+G(m_Q^2-2)}{(1+3G)m_Q^2}&0&0&\frac{2\sqrt{2}G\xi}{m_{Q}\Lambda\sqrt{1+3G}}&\frac{2\sqrt{2}G\xi}{m_Q^2\Lambda\sqrt{1+3G}}&0\\
        
        0&0&0&1&0&0\\
        
        0&-\frac{2\sqrt{2}G\xi}{m_{Q}\Lambda\sqrt{1+3G}}&-1+\frac{4G\xi^2}{m_Q^2\Lambda^2}&0&0&0\\
        
        0&0&0&0&0&1\\
        
        \frac{2\sqrt{2}G\xi}{m_Q^2\Lambda\sqrt{1+3G}}&0&0&0&-1+\frac{4G\xi^2}{m_Q^2\Lambda^2}&0
    \end{pmatrix}\,. 
\end{equation}
The most general solution is given by a linear combination of the eigenvalues $w_i$ and the associated eigenvectors $\bold{s}_i$ of Eq.~\eqref{Coefficienteq} in the form\footnote{Notice that here we are assuming all the components of $\mathcal{N}$ to be constant in the infinite past.} 

\begin{equation}\label{DefinitionD}
    \mathcal{R}_{\rm in}=\sum_{i=1}^{6} c_{i}\bold{s}_{i}e^{w_{i}x}\bigg|_{x=x_{\rm in}}\,,
\end{equation}
where the $c_{i}$'s are the integration constants accompanying every term in the solution. The eigenvalues $w_{i}$ have the following form:
\begin{equation}
\begin{split}\label{eigenv}
    \hspace{-5mm}w=\Bigg\{-i,i,-\frac{i\sqrt{m_{Q}^{2}\Lambda^{2}-4G\xi^{2}}}{m_{Q}\Lambda},&\frac{i\sqrt{m_{Q}^{2}\Lambda^{2}-4G\xi^{2}}}{m_{Q}\Lambda},-\frac{i\sqrt{m_Q^2\Lambda^2+Gm_Q^2\Lambda^2-2G\Lambda^2-4G\xi^2-4G^2\xi^2}}{m_{Q}\Lambda\sqrt{1+3G}},\\
    &\frac{i\sqrt{m_Q^2\Lambda^2+Gm_Q^2\Lambda^2-2G\Lambda^2-4G\xi^2-4G^2\xi^2}}{m_{Q}\Lambda\sqrt{1+3G}}\Bigg\}\,.
\end{split}
\end{equation}
By inspection of these eigenvalues one finds that, for the solutions to be stable, the following conditions must be in place:
\begin{equation}\label{Stabilityconditions}
\begin{aligned}
    &m_Q^2\Lambda^2+Gm_Q^2\Lambda^2-2G\Lambda^2-4G\xi^2-4G^2\xi^2>0\,, \\
    &m_Q^2\Lambda^2-4G\xi^2>0\,.
\end{aligned}
\end{equation}
which can be rewritten in terms of the model parameters as
\begin{equation} \label{cond}
\begin{aligned}
    &(1+g^2\lambda_1Q^4)(1-\lambda_1\dot{\chi}^2)-2\lambda_1H^2Q^2>0\,, \\
    &1-\lambda_1\dot{\chi}^2>0\,,
\end{aligned}
\end{equation}
where Eq.~(\ref{par}) was used. The second condition in Eq.~(\ref{cond}) is nothing but the positivity requirement for the tensor sound speed, from which we had previously derived a lower bound on $\lambda$ (see Eq.~\eqref{ConditionOnLambda}). Another, slightly stricter, condition on the same parameter follows from the first line of Eq.~(\ref{cond}), which is equivalent to requiring
\begin{equation}\label{lambdaLowerBound}
    \lambda\gtrsim\frac{6f\sqrt{\lambda_1}H^2}{gQ}\,.
\end{equation}
This relation ensures that all the eigenvelues are such that ${\rm Re}[w_i]=0$, three with ${\rm Im}[w_i]>0$ (positive frequency) and three  with ${\rm Im}[w_i]<0$ (negative frequency). For the sake of simplicity, we anticipate here that imposing the generalized Wronskian conditions enforces the physical solutions to be the ones with positive frequency, so that we set $c_{1}=c_{3}=c_{5}=0$. \\
\indent Moving on, since we are interested in the solution for the fields (and not their derivatives), we will focus solely on the first, third and fifth entries of Eq.~\eqref{higherdimsystem}, for which the most general solution  reads
\begin{equation}\label{MostGeneralD}
\begin{split}
    \Delta_{\rm in}&=c_{1}\bold{s}_{1}\exp(ix)+c_{2}\bold{s}_{2}\exp\left(\frac{ix\sqrt{m_Q^2\Lambda^2-4G\xi^2}}{m_{Q}\Lambda}\right)\\
    &+c_{3}\bold{s}_{3}\exp\left(\frac{ix\sqrt{m_Q^2\Lambda^2+Gm_Q^2\Lambda^2-2G\Lambda^2-4G\xi^2-4G^2\xi^2}}{m_{Q}\Lambda\sqrt{1+3G}}\right)\Bigg|_{x=x_{\rm in}}\,,
\end{split}
\end{equation}
where we have renamed the labels for the integration constants and the eigenvectors accordingly. The latter are given by: 
\begin{equation}
    \bold{s}_{1}=\begin{pmatrix}
        -\frac{2i\xi\sqrt{1+3G}}{\Lambda\sqrt{2}}\\ \\
        -m_{Q}\\\\ 
        i
    \end{pmatrix}, \quad \bold{s}_{2}=\begin{pmatrix}
        0\\\\
        \frac{m_{Q}\Lambda^{2}}{m_Q^2\Lambda^2-4G\xi^2}\\\\
        \frac{im_{Q}\Lambda}{\sqrt{m_Q^2\Lambda^2-4G\xi^2}}
    \end{pmatrix}, \quad \bold{s}_{3}=\begin{pmatrix}
        \frac{i\left(\left(1+m_{Q}^{2}\right)\Lambda^{2}m_{Q}-4G\xi m_{Q}\right)}{\xi\sqrt{2}\sqrt{m_Q^2\Lambda^2+Gm_Q^2\Lambda^2-2G\Lambda^2-4G\xi^2-4G^2\xi^2}}\\\\
        -m_{Q}\\\\
        \frac{im_{Q}\Lambda\sqrt{1+3G}}{\sqrt{m_Q^2\Lambda^2+Gm_Q^2\Lambda^2-2G\Lambda^2-4G\xi^2-4G^2\xi^2}}
    \end{pmatrix}\,.
\end{equation}
Promoting the classical fields to quantum operators ($\Delta_i\rightarrow\mathcal{D}_{ij}$, see Eq.~\eqref{DeltaDecom}) one then writes 
\begin{equation}\label{MostGeneralD_{ij}}
\begin{split}
    \mathcal{D}_{ij, \rm in }&=c_{1j}s_{1i}\exp(ix)+c_{2j}s_{2i}\exp\left(\frac{ix\sqrt{m_Q^2\Lambda^2-4G\xi^2}}{m_{Q}\Lambda}\right)\\
    &+c_{3j}s_{3i}\exp\left(\frac{ix\sqrt{m_Q^2\Lambda^2+Gm_Q^2\Lambda^2-2G\Lambda^2-4G\xi^2-4G^2\xi^2}}{m_{Q}\Lambda\sqrt{1+3G}}\right)\Bigg|_{x=x_{\rm in}}\,,
\end{split}
\end{equation}
where now $j\in (1,3)$ correspond to the associated basis of creation-annihilation operators in Eq.~\eqref{DeltaDecom}. \\
\indent Eq.~\eqref{MostGeneralD_{ij}} contains eighteen real constants (corresponding to the nine complex arbitrary constants $c_{ij}$). Nine of the real constants can be fixed using the generalized Wronskian conditions, Eqs.~\eqref{MultifieldWronsk}-\eqref{MultifieldWronsk2}. The remaining nine encode the freedom intrinsic to our multifield system, following from the invariance of the physical observables under a transformation $\mathcal{D}\rightarrow \mathcal{D}\cdot U$ ($U$ being a unitary matrix). The latter is entirely analogous to the phase arbitrariness of the mode function in the single-field case. 
In this regard, a simple choice is to set $c_{ij}=0$ for $i\neq j$ and ${\rm Im}[c_{ii}]=0$. The Wronskian conditions then lead to the following expressions for the remaining coefficients 
\begin{subequations}
\begin{align}
    &\left(\sqrt{2k}c_{11}\right)^{2}=\frac{\Lambda^2}{(1+m_Q^2)\Lambda^2+2(1+G)\xi^2}\,, \\    &\left(\sqrt{2k}c_{22}\right)^{2}=\frac{(m_Q^2\Lambda^2-4G\xi^2)^{3/2}}{m_Q(1+m_Q^2)\Lambda^3-4Gm_Q\Lambda\xi^2}\,,\\
    &\left(\sqrt{2k}c_{33}\right)^{2}=\frac{2\Lambda\xi^2\sqrt{(1+3G)((m_Q^2+G(-2+m_Q^2))\Lambda^2-4G(1+G)\xi^2)}}{m_Q((1+m_Q^2)\Lambda^2-4G\xi^2)((1+m_Q^2)\Lambda^2+2(1+G)\xi^2)}\,.
\end{align}
\end{subequations}
In order to write down the initial conditions for the full scalar system, the last step is to input these integration constants into the solution in Eq.~\eqref{MostGeneralD_{ij}}. 

\indent Analogously to the tensor case, the dimensionless power spectra for scalar perturbations can be computed as
\begin{equation}\label{scalarspectra}
    \mathcal{P}^{(S)}_{ij}=\frac{k^3}{2\pi^2}{\rm Re}\left[\mathcal{M}_{\Delta,ik}\mathcal{M}_{\Delta,j\ell}\mathcal{D}_{km}\mathcal{D}^*_{\ell m}\right]\,.
\end{equation}
We are now ready to numerically evolve the full equations for the scalar sector.

\subsection{Numerical evolution}

\begin{figure}[t!]
    \includegraphics[scale=0.35]{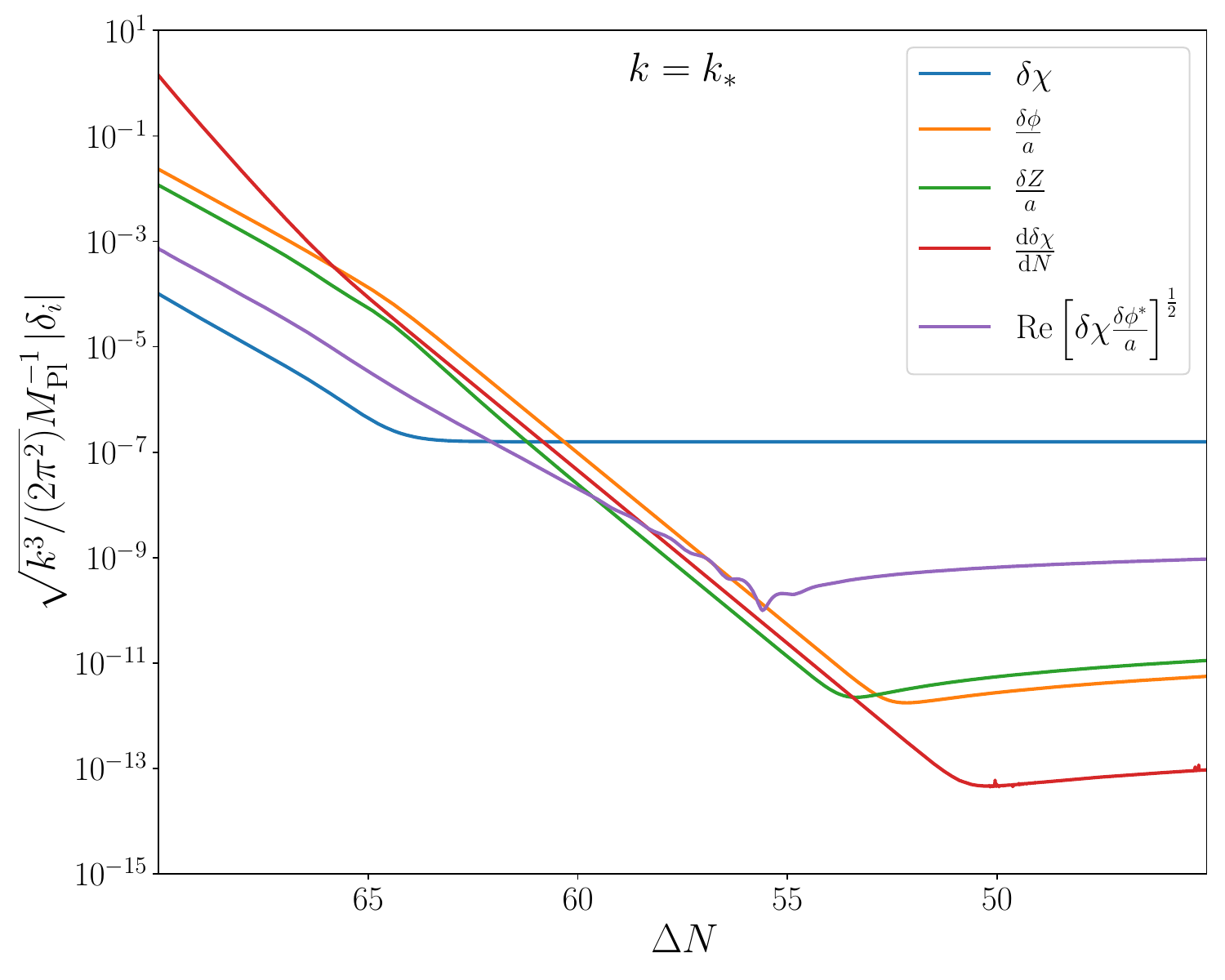}
    \caption{\textit{Plot of the main contributions to the two-point function of the full expression for $\zeta$ (Eq.~(\ref{ZetaFull})), considering the fiducial parameters from Eq.~(\ref{benchmark2}). From this plot it is clear that the main contribution to the curvature  power spectrum comes from the scalar mode $\delta\chi$.}}\label{Correlators}
\end{figure}

Let us now compute the (gauge-invariant) curvature perturbation $\zeta$, defined as
\begin{equation}\label{zetaDef}
    \zeta\equiv-\frac{H}{\dot{\rho}}\delta\rho^{\rm flat}\,.
\end{equation}
The energy density perturbation in the spatially flat gauge, $\delta\rho^{\rm flat}$, is obtained by expanding the total stress-energy tensor $T_{\mu\nu}$ at linear order, and using the standard relations $\rho=T^0_{ \ 0}$, $P=\frac{1}{3}T^i_{ \ i}$ alongside the continuity equation: 
\begin{equation}\label{Continuity}
    \dot{\rho}+3H(\rho+P)=0\,.
\end{equation}
Therefore, $\zeta$ takes the following form:
\begin{equation}\label{ZetaFull}
    \zeta=\frac{1}{6\epsilon_HH^2\Mp^2}\left(V'(\chi)\delta \chi+\dot{\chi}\delta \dot{\chi}(1+3 g^2\lambda_1 Q^4)+6g^2Q^3(1+\lambda_1\dot{\chi}^2)\frac{\delta\phi}{a}+3(\dot{Q}+HQ)\frac{\delta\dot{\phi}}{a}\right)\,.
\end{equation}
Employing the background EoM (Eq.~\eqref{SlowRollPot}), it is straightforward to verify that the first term in $\zeta$ is equal to the one found in single-field slow-roll inflation, namely $\zeta\simeq-\frac{H}{\dot{\chi}}\delta\chi$. One can also check that the remaining terms in Eq.~\eqref{ZetaFull} are sub-leading (see Fig.~\ref{Correlators}).
With this in mind, the (dimensionless) curvature power spectrum is computed as

\begin{equation}\label{zetaspectrum}
    \mathcal{P}_{\zeta}\simeq\frac{k^3}{2\pi^2}\frac{H^{2}}{a^2\dot{\chi}^{2}(1+3g^{2}\lambda_{1}Q^{4})}\sum_{m=1}^3|\mathcal{D}_{1m}|^2\,,
\end{equation}
where we used Eq.~\eqref{scalarspectra} with $i=j=1$. One can verify that this result and the power spectrum obtained with the full expression for $\zeta$ in Eq.~\eqref{ZetaFull} are  equivalent on super-horizon scales, thus justifying the use  of Eq.~\eqref{zetaspectrum} in computing  late-time observables. \\

\subsection{Observables}\label{Observables}

Throughout this work, we have employed the fiducial set of parameters  in Eq.~(\ref{benchmark2}). These  have been chosen as they fulfill a number of necessary requirements. First of all, we restrict ourselves to the weak coupling regime $g<1$. We keep $f$ sub-Planckian and limit the range for $\mu$ such that the tensor-to-scalar ratio remains below the experimental upper bound. Next, using Eq.~\eqref{Nfbounds}, we enforce a duration of inflation of at least sixty $e$-folds and obtain a lower bound on the kinetic gauge coupling $\lambda_{1}$. As for the Chern-Simons coupling  $\lambda$, we enforce the lower bound in Eq.~\eqref{lambdaLowerBound}, derived from the stability conditions on scalar perturbations. This will automatically  ensure also the positivity of the tensor sound speed $c_{t}$. At the same time, we consider $\lambda$ values  compatible with the regime where the CS's effect on the background remains negligible compared to kinetic gauge friction. \\
\indent Having narrowed down the parameter space, we searched for a set that is  viable in view of the  constraints on $A_{s}$, $n_{s}$ and $r$. We first identify the end of inflation by numerically evolving the background system until the time when $\epsilon_{H}=1$. We then count backwards in order to determine the horizon-crossing of the CMB scale $k_*$, which we conventionally set to be at sixty $e$-folds before the end of inflation. Finally, we evolve the scalar and tensor modes with $k=k_*$ (or $k=k_*/25$, in the tensor-to-scalar ratio), in order to obtain predictions for the observables that can be compared with current CMB constraints \cite{BICEP:2021xfz}. These are defined as
\begin{equation}
    \mathcal{P}_\zeta=A_s\left(\frac{k}{k_*}\right)^{n_s-1}\,, \quad r=\frac{\mathcal{P}_h^{\rm tot}}{\mathcal{P}_\zeta}\,.
\end{equation}

\noindent As our work is the first step into the study of the kinetic gauge coupling,  we did not aim at an exhaustive exploration of the parameter space but rather focused on identifying viable values of the parameters
vis-\`{a}-vis CMB observations at the $2\sigma$ level. For our fiducial set of parameters we obtain:

\begin{center}
\begin{tabular}{||c c c||} 
 \hline
 $A_s$ & $n_{s}$ & $r$  \\ [0.5ex] 
 \hline\hline
 $2\times10^{-9}$ & $0.96$ & $4.6\times 10^{-3}$ \\ [1ex] 
 \hline
\end{tabular}
\end{center}
We stress here that as one move towards smaller and smaller scales, the particle production parameter $m_Q$ will grow and so will the (enhanced polarization of)  tensor fluctuations in the gauge sector. Such enhancement may well become so large that its backreaction on the background equations can no longer be ignored. We leave a thorough exploration of strong backreaction dynamics to future work and only offer some preliminary comments in Appendix \ref{note}.

\section{Conclusions}\label{Conclusions}

In this work we introduced a model with kinetic gauge friction in chromo-natural (KFC) inflation. Natural inflation and its extensions have been extensively studied in the literature, given their ability to tackle the inflationary \textit{eta problem}. Considering the presence of gauge sectors next to ALPs during inflation has proven fruitful in that (i) gauge fields do not spoil the shift symmetry while (ii) delivering an interesting GW phenomenology  and (iii) a possible mechanism for re-heating. The novelty in our work lies in the introduction of kinetic coupling between the gauge and axion-inflaton sector. We have shown how such coupling can provide an important source of friction, supporting an inflationary phase of over sixty $e$-folds for sub-Planckian values of the ALP decay constant $f$. The perturbative analysis revealed a negative sound speed, whose sign we are able to flip upon considering a Chern-Simons coupling between gauge and axion sectors.~The background dynamics remains largely driven by the kinetic coupling and is therefore rather different from that, for example, of the standard chromo-natural inflation model.~Nevertheless, the presence of the CS term is crucial for the stability of perturbations.

In terms of observables, we identify a region of parameter space that is compatible with current CMB bounds. The gravitational wave spectrum may be chiral in light of the enhancement of one polarization of the tensor degrees of freedoms in the gauge sector.  Linearly coupled to tensor metric fluctuations,  whenever such mode provides the leading contribution to gravitational waves the signal acquires a non-trivial chirality.

In order to extend our analysis and consider the gravitational signal at intermediate and small scales one should take into account the strong backreaction regime\footnote{This notion is based on the well-tested expectation that the damping of energy from the ALP to the gauge sector will increase as the inflaton velocity grows.}. Growing fluctuations in the gauge sector can significantly affect background dynamics and alter predictions for cosmological observables. One can qualitatively tackle the strong backreaction regime via numerical methods, whilst lattice simulations remain the gold standard. We plan to take on backreaction in a forthcoming work. Another related aspect worth investigating is perturbativity bounds. A description of KFC inflation valid in the perturbatively accessible region of the strong backreaction regime will enable the use of current and upcoming gravitational wave probes (such as LISA, Einstein Telescope, Cosmic Explorer, and DECIGO) to test the parameter space of the model. 

It would be interesting to explore the model in different directions, including its reduction to a non-standard single-field scenario upon integrating out gauge fields.~Another aspect worthy of  investigation is the existence of string theory realizations of KFC inflation and the top-down constraints these might enforce on its parameter space. \\
\vspace{1cm}


\noindent{\bf Acknowledgments.} RGQ thanks CONAHCyT and RUG for funding and support. MF acknowledges support from the “Consolidaci\'{o}n Investigadora” grant CNS2022-135590 and the “Ram\'{o}n y Cajal” grant RYC2021-033786-I. MF's work is partially supported by the Spanish Research Agency (Agencia Estatal de Investigaci\'{o}n) through the Grant IFT Centro de Excelencia Severo Ochoa No CEX2020-001007-S, funded by MCIN/AEI/10.13039/501100011033.

\newpage
\appendix

\section{General system of background equations }\label{appA}
In this appendix we present the most general EoM derived from Eq.~\eqref{GaugeAction2}:

\begin{align}\label{Friedmaneq}
     &3\Mp^2H^2=\frac{1}{2} \dot{\chi}^2 \Big[1+6g^{2}\lambda_{2}Q^{4}+( \lambda_{1}-2\lambda_{2})\left(18 HQ\dot{Q}+9H^{2} Q^{2}+9 \dot{Q}^{2}\right)\Big]\, ,\\
   \nonumber\\
    &-2\Mp^2\dot{H}=\dot{\chi}^{2}\left[1+2 g^{2} \lambda_{2}Q^{4}+( \lambda_{1}-2 \lambda_{2}) \left(10HQ \dot{Q}+5H^{2} Q^{2}+5\dot{Q}^{2}\right)\right]+2 \left(H Q+\dot{Q}\right)^{2}+2 g^{2} Q^{4}\, ,\\
   &\ddot{\chi}\left[1+6 g^{2}\lambda_{2} Q^{4}+(\lambda_{1}-2\lambda_{2}) \left(6 HQ\dot{Q}+3H^{2}Q^{2}+3 \dot{Q}^{2}\right)\right] \\
    &+3H\dot{\chi} \Bigg[1+(\lambda_{1}-2 \lambda_{2}) \left(\frac{2Q \dot{H}\dot{Q}}{H}+2Q^{2}\dot{H}+8 HQ\dot{Q}+\frac{2 \dot{Q}\ddot{Q}}{H}+3H^{2}Q^{2}+2 Q\ddot{Q}+5 \dot{Q}^2\right)\nonumber\\
    &+\frac{8g^{2}\lambda_{2}Q^{3} \dot{Q}}{H}+6g^{2} \lambda_{2}Q^4\Bigg]+V'(\chi)=0\, ,\nonumber\\
    \nonumber\\
    &\ddot{Q} \left[1+(\lambda_{1}-2 \lambda_{2}) \dot{\chi}^{2}\right]+3H\dot{Q}\left[1+(\lambda _{1}-2\lambda_{2}) \left(\frac{2 \dot{\chi}\ddot{\chi}}{3H}+\dot{\chi}^{2}\right)\right]+2g^{2}Q^{3}-4g^{2}\lambda_{2} Q^{3}\dot{\chi}^{2}\\
    &+Q\left[\dot{H}+2 H^{2}+\left(\lambda_{1}-2\lambda_{2}\right) \left(\dot{H}\dot{\chi}^{2}+2H^{2}\dot{\chi}^{2}+2H\dot{\chi}\ddot{\chi}\right)\right]=0\, .\nonumber
\end{align}
It is clear that one can explore different combinations of $\lambda_{1}$ and $\lambda_{2}$ to study the effects of specific contributions to the dynamics of the model. The case considered in the main text, with $\lambda_{2}=\lambda_{1}/2$, is special as not only does it reduce the complexity of the system but at the same time it cancels out all the Abelian terms in the kinetic gauge friction. Therefore, only terms proportional to $g^{2}$ are born from this new coupling.
\section{Full expressions for scalar mode matrices}
\label{app:B}
We report here the complete expressions for the entries of the matrices appearing in the EoM for scalar perturbations (Eq.~\eqref{ScalarEoM}):
\begin{align}\label{alphadeltafull}
    \alpha_{\Delta,12} &=-\alpha_{\Delta,21}=-\frac{\sqrt{2} x}{\sqrt{x^2+2 m_Q^2}\sqrt{1+3 G}} \left[\frac{2 G \xi}{\Lambda  m_Q}\left(1+\frac{4 m_Q^2}{x^2}\right)+\frac{\Lambda  m_Q^2}{x^2}\right]\, ,\\
    \alpha_{\Delta,13} &=-\alpha_{\Delta,31}=\frac{\sqrt{2} \Lambda  m_Q}{\sqrt{1+3 G}x}\left(1+\frac{8 G \xi }{\Lambda ^2 m_Q}\right)\, ,
\end{align}

\begin{align}\label{betadeltafull}
    \beta_{\Delta,11} &=\frac{1}{(1+3G)x^2}\Bigg\{x^2 \left(1+G-\frac{2 G}{m_Q^2}\right)-(2-\epsilon_H)(1-3 G)+12 G \left[m_Q^2 \left(1-\frac{4 G \xi ^2}{\Lambda ^2 m_Q^2}\right)-\xi  m_Q\right]\nonumber \\
    &+\frac{x^2}{x^2+2m_Q^2}\left(\Lambda ^2 m_Q^2+8 G \xi  m_Q+\frac{16 G^2 \xi ^2}{\Lambda ^2}\right)+\frac{V''(\chi)}{H^2}\Bigg\}\, , \\\nonumber\\
    \beta_{\Delta,12} &=\frac{x^2}{(x^2+2m_Q^2)^{3/2}(1+3 G)^{3/2}}\Bigg\{\sqrt{2}\Bigg[ \Lambda  \left(1+\frac{4 m_Q^4}{x^4}\right)(1-3 G)-\frac{6 G \xi }{\Lambda  m_Q}\left(1+\frac{4 m_Q^2}{x^2}+\frac{8 m_Q^4}{3 x^4}\right)(1+3 G) \\
    &+\frac{3 \Lambda  m_Q^2}{x^2}(1-5 G)\Bigg]+\frac{G\sqrt{\epsilon_H}}{\sqrt{\epsilon_Q}}\left(1+\frac{4 m_Q^2}{x^2}+\frac{4 m_Q^4}{x^4}\right)\left[-12+\epsilon_Q \left(6+\frac{6}{m_Q^2}+\frac{8 \xi ^2}{\Lambda ^2 m_Q^2}(1+3 G) \right)\right]\Bigg\}\, \nonumber, \\\nonumber\\
    \beta_{\Delta,13} &= \frac{1}{(1+3 G)^{3/2}}\Bigg\{-\frac{2 \sqrt{2} G \xi}{\Lambda  m_Q^2}\left(1-\frac{4 m_Q^2}{x^2}\right)(1+3 G)-\frac{2 \sqrt{2} \Lambda  m_Q}{x^2}(1-3 G) \\
    &+\frac{G \sqrt{\epsilon_H}}{\sqrt{\epsilon_Q}x^2 }\left[12 m_Q \left(2-\epsilon_Q-\frac{\epsilon_Q}{m_Q^2}\right)-\frac{16 \xi ^2 \epsilon_Q}{\Lambda ^2 m_Q}(1+3 G)\right]\Bigg\}\, , \nonumber\\\nonumber\\
    \beta_{\Delta,21} &=\frac{1}{(x^2+2m_Q^2)^{3/2}(1+3 G)^{3/2} }\Bigg\{\sqrt{2} \left[\Lambda  x^2+\Lambda  m_Q^2(1-3 G) +\frac{2 \Lambda  m_Q^4}{x^2}(1+3 G)\left(1+\frac{8 G \xi }{\Lambda ^2 m_Q}\right)\right] \\
    &+\frac{G\sqrt{\epsilon_H}}{\sqrt{\epsilon_Q}}\left[-6( x^2+2m_Q^2)+\epsilon_Q \left(3 x^2 \left(1+\frac{1}{m_Q^2}\right)+6 (1+m_Q^2)+\frac{4 \xi ^2}{\Lambda ^2 m_Q^2}(x^2+2m_Q^2)(1+3 G)\right)\right]\Bigg\}\, , \nonumber\\\nonumber\\
    \beta_{\Delta,22} &=\left(1+\frac{2 m_Q^2}{x^2}\right)\left(1-\frac{4 G \xi ^2}{\Lambda ^2 m_Q^2}\right)+\frac{2}{x^2+2m_Q^2} \left[m_Q^2 \left(1-\frac{4 G \xi ^2}{\Lambda ^2 m_Q^2}\right)-\xi  m_Q+\frac{3 m_Q^2}{x^2+2m_Q^2}\right]\, , \\
    \beta_{\Delta,23} &=\beta_{\Delta,32}=\frac{2\sqrt{x^2+2m_Q^2}}{x^2}\left[\xi-m_Q\left(1-\frac{4 G \xi ^2 }{\Lambda ^2 m_Q^2}\right)\right]\, , \\
    \beta_{\Delta,31} &=-\frac{\sqrt{2}}{\sqrt{1+3G}}\left(\frac{2 G \xi }{\Lambda  m_Q^2}+\frac{\Lambda  m_Q}{x^2}+\frac{8 G \xi }{\Lambda  x^2}\right)\, , \\
    \beta_{\Delta,33} &=\left(1+\frac{4 m_Q^2}{x^2}\right)\left(1-\frac{4 G \xi ^2}{\Lambda ^2 m_Q^2}\right)-\frac{2 \xi  m_Q}{x^2}\, . 
\end{align}
It is straightforward to verify that the CNI limit is obtained by setting $G=0$.


\section{A brief note on backreaction}
\label{note}

In this work we have neglected the effect that the enhanced polarization of the tensor mode $T_R$ might have on the background dynamics. We can obtain a first estimate by including the backreaction terms in the background EoM as follows:
\begin{equation}
\begin{aligned}
    &3\Mp^2H^2=\frac{1}{2}\dot{\chi}^2+V(\chi)+\frac{3}{2}\left(\dot{Q}+HQ\right)^2+\frac{3}{2}g^2Q^4+\frac{3}{2}g^2\lambda_1Q^4\dot{\chi}^2+\rho_t\,, \\
    &\left(1+3g^2\lambda_1Q^4\right)\ddot{\chi}+3\left(H+g^2\lambda_1Q^3\left(4\dot{Q}+3HQ\right)\right)\dot{\chi}+V'(\chi)=-\frac{3g\lambda}{f}Q^2\left(\dot{Q}+HQ\right)+\mathcal{T}^\chi_{\rm BQ}\,, \\
    &\ddot{Q}+3H\dot{Q}+\left(\dot{H}+2H^2\right)Q+2g^2Q^3=\frac{g\lambda}{f}\dot{\chi}Q^2+2g^2\lambda_1Q^3\dot{\chi}^2+\mathcal{T}^Q_{\rm BQ}\,,
\end{aligned}
\end{equation}
with
\begin{equation}
\begin{aligned}
     \rho_t&=\frac{1}{2a^2}\int\frac{{\rm d}^3 k}{(2\pi^3)}\left[\left|\frac{{\rm d}}{{\rm d}t}T_R\right|^2+(1+\lambda_1\dot{\chi}^2)\left(\frac{k^2}{a^2}-2gQ\frac{k}{a}\right)|T_R|^2\right]\,,\\
    \mathcal{T}^\chi_{\rm BQ} &=\frac{1}{a^3}\frac{{\rm d}}{{\rm d}t}\int\frac{{\rm d}^3 k}{(2\pi^3)}\left(\frac{\lambda gaQ}{2f}-\frac{\lambda k}{2f}-\frac{k^2\lambda_1\dot{\chi}}{a}+2gk\lambda_1Q\dot{\chi}\right)|T_R|^2\,,\\
    \mathcal{T}^Q_{\rm BQ}&=-\frac{g}{3a^2}\int\frac{{\rm d}^3 k}{(2\pi^3)}\left(\frac{\lambda\dot{\chi}}{2f}-\frac{k}{a}(1-\lambda_1\dot{\chi}^2)\right)|T_R|^2\,.
\end{aligned}
\end{equation}
One can  estimate the backreaction terms by (see \textit{e.g.} \cite{Dimastrogiovanni:2016fuu}) making use of the (non-backreacting) background Eqs.~\eqref{KGFXEoM2}-\eqref{KGFQEoM2}. For example, one may use the solution in Eq.~\eqref{TRanalytical} for $T_R$ and numerically sample the $k$-integrals at different times during inflation. We leave the analysis of the strong backreaction regime to future work.


\bibliographystyle{JHEP}
\bibliography{biblio}

\end{document}